\documentclass[11pt,fleqn,a4paper]{article}
\usepackage{amsfonts,amsmath,amssymb,amsbsy,color,epsfig,harvard,multirow,pifont,rotating,threeparttable}
\usepackage{accents}
\begin{document}
\bibliographystyle{joe}
\citationstyle{dcu}
\newcommand{\cn}{\citeasnoun}
\newcommand{\cnm}{\citename}
\newcommand{\pc}{\possessivecite}
\newcommand{\ca}{\citeaffixed}
\newcommand{\cy}{\citeyear*}
\newcommand{\E}{\mathbb{E}}
\newcommand{\V}{\mathbb{V}}
\newcommand{\Prob}{\mathrm{Pr}}
\newcommand{\calE}{\mathcal{E}_n(r)}
\newcommand{\calEone}{\mathcal{E}_n^1(r)}
\newcommand{\calEtwo}{\mathcal{E}_n^2(r)}
\newcommand{\calEthree}{\mathcal{E}_n^3(\epsilon)}
\newcommand{\calEc}{\breve{\mathcal{E}}_n(r)}
\newcommand{\calEonec}{\breve{\mathcal{E}}_n^1(r)}
\newcommand{\calEtwoc}{\breve{\mathcal{E}}_n^2(r)}
\newcommand{\calEthreec}{\breve{\mathcal{E}}_n^3(\epsilon)}
\newcommand{\floor}[1]{\lfloor{#1}\rfloor}
\newcommand{\norm}[1]{\left\Vert{#1}\right\Vert}
\newcommand{\abs}[1]{\left\vert{#1}\right\vert}
\newcommand{\given}{\,|\,}
\newcommand{\bs}{\boldsymbol}
\newcommand\ubar[1]{\underaccent{\bar}{#1}}
\newcommand{\AMSE}{\mathrm{AMSE}}
\newcommand{\MSE}{\mathrm{MSE}}
\newcommand{\trace}{\mathrm{tr}}
\newcommand{\supp}{\mathrm{supp}}
\newcommand{\Null}{\mathrm{null}}
\lineskip=.35ex \baselineskip 3.5ex \mathindent 0em
\setlength{\arraycolsep}{0.02in}
\begin{center}
\noindent\textbf{\Large Smoothing quantile regressions}\vskip 2em
\begin{tabular}{ccc}
\textbf{Marcelo Fernandes}          && \textbf{Emmanuel Guerre}\\
Sao Paulo School of Economics, FGV  && University of Kent\\
\texttt{marcelo.fernandes@fgv.br}   && \texttt{e.o.guerre@kent.ac.uk}\\\\
\multicolumn{3}{c}{\textbf{Eduardo Horta}}\\
\multicolumn{3}{c}{Universidade Federal do Rio Grande do Sul}\\
\multicolumn{3}{c}{\texttt{eduardo.horta@ufrgs.br}}
\end{tabular}
\vskip 2em
This version: \today
\end{center}\vskip 2em
\noindent\textbf{Abstract}~~~~We propose to smooth the objective function, rather than only the indicator on the check function, in a linear quantile regression context. Not only does the resulting smoothed quantile regression estimator yield a lower mean squared error and a more accurate Bahadur-Kiefer representation than the standard estimator, but it is also asymptotically differentiable. We exploit the latter to propose a quantile density estimator that does not suffer from the curse of dimensionality. This means estimating the conditional density function without worrying about the dimension of the covariate vector. It also allows for two-stage efficient quantile regression estimation. Our asymptotic theory holds uniformly with respect to the bandwidth and quantile level. Finally, we propose a rule of thumb for choosing the smoothing bandwidth that should approximate well the optimal bandwidth. Simulations confirm that our smoothed quantile regression estimator indeed performs very well in finite samples.\vfill
\begin{small}
\noindent\textbf{JEL classification numbers}~~~~C14, C21\\
\textbf{Keywords}~~~~asymptotic expansion, Bahadur-Kiefer representation, conditional quantile, convolution-based smoothing, data-driven bandwidth.
\end{small}\vfill

\lineskip=.25ex \baselineskip 2.5ex
\begin{footnotesize}
\noindent\textbf{Acknowledgments}~~~~We are grateful to Antonio Galv\~{a}o and Aureo de Paula for valuable comments, as well as to the Associate Editor and anonymous referees. Fernandes also thanks the financial support from CNPq (302278/2018-4).
\end{footnotesize}

\newpage\pagestyle{plain}
\lineskip=.5ex \baselineskip 5ex
\section{Introduction}

Quantile regression (QR) enjoys some very appealing features. Apart from enabling some very flexible patterns of partial effects, quantile regressions are also interesting because they satisfy some equivariance and robustness principles. See \cn{koenker1978regression} and \cn{koenker2005quantile} for theoretical aspects; and \cn{koenker2000galton}, \cn{buchinsky1998recent}, \cn{koenker2001quantile}, \cn{koenker2005quantile} and references therein for applications.

There is a price to pay, though. The objective function that the standard QR estimator aims to minimize is not smooth. As established in \cn{bassett1982empirical}, it follows that the paths of this estimator have jumps, even if the underlying quantile function is very regular. Because the objective function does not have second derivatives, statistical inference is not straightforward and involves ancillary estimation of nuisance parameters (namely, the asymptotic covariance matrix depends on the population conditional density evaluated at the true quantile). See the discussions in \cn{koenker1994confidence}, \cn{buchinsky1995estimating}, \cn{koenker2005quantile}, \cn{goh2009nonstandard}, and \cn{fan2016direct}, among others. Unsurprisingly, the literature now boasts a wide array of techniques to tackle this issue, including bootstrapping \cite{horowitz1998bootstrap,machado2005bootstrap}, MCMC methods \cite{chernozhukov2003mcmc}, empirical likelihood \cite{otsu2008conditional,whang2006smoothed}, strong approximation methods \cite{portnoy2012nearly}, nonstandard inference \cite{goh2009nonstandard}, and other nonparametric approaches \cite{mammen2017expansion,fan2016direct}.

Further, the asymptotic normality of the standard QR estimator relies on Bahadur-Kiefer representations with poor rates of convergence. The latter is at best of order $n^{-1/4}$ for iid (homoskedastic) errors \ca{koenker1987estimation,knight2001comparing,jurevckova2012methodology}{see}. This means that the first-order \emph{linear} Gaussian approximation for the distribution of the QR estimators could well fail in finite samples. \cn{portnoy2012nearly} however obtains a nearly $n^{-1/2}$ rate using a nonlinear approximation, explaining perhaps why QR inference is actually rather good in finite samples despite the poor rates of the Bahadur-Kiefer remainder.

In this paper, we employ a convolution-type smoothing of the objective function that helps produce a continuous QR estimator, which is not only less irregular and variable than the standard estimator but also more linear, in the sense that the stochastic order of the Bahadur-Kiefer remainder term is much closer to $n^{-1/2}$ for a proper bandwidth choice. As the smoothing we propose ensures the twice differentiability of the objective function with respect to the parameter vector, we can readily estimate the asymptotic covariance matrix of our smoothed QR estimator using a standard sandwich formula. In fact, the second derivative of our kernel-smoothed objective function indeed coincides with the usual kernel-based covariance matrix estimators in the standard QR setup.

Quantile estimation aims to offer a global picture of the distribution. Bearing this in mind, we contemplate an asymptotic theory that holds uniformly with respect to quantile levels, or in a functional sense. This is in line with the functional central limit theorem that \cn{koenker2002inference} establish for specification-testing purposes. Apart from uniformity in quantile level, we also focus on bandwidth uniformity in consistency and asymptotic distribution results in order to provide some robustness against bandwidth snooping. Practical applications of smoothing procedures may indeed involve several bandwidth choices, exposing practitioners to the risk of biasing inference by selecting a bandwidth that agrees with prior beliefs. As a result, our asymptotic theory considers uniformity in both quantile level and bandwidth, allowing for data-driven choices for the latter. This is in sharp contrast with related works on smoothing, such as \cn{horowitz1998bootstrap} and \cn{kaplan2017smoothed}. In particular, they restrict attention to a given quantile level and to deterministic bandwidths when studying higher-order accuracy for confidence interval coverage and type-I test errors, respectively.

Our main contributions are as follows. First, smoothing the QR objective function induces bias in the QR estimation. We derive the order of such bias, showing that it is negligible with respect to $\sqrt{n}$ for the optimal bandwidth rate, as well as for the data-driven bandwidth we use in the Monte Carlo experiments. Second, we establish that the smoothed QR estimator has a smaller Bahadur-Kiefer linearization error than the standard QR estimator as long as the bandwidth remains large enough. Third, we find that, for a proper bandwidth choice, our smoothed QR estimator yields a lower asymptotic mean squared error (AMSE) than the standard estimator. Moreover, we derive the optimal bandwidth for our smoothed QR estimator. Fourth, we provide a covariance matrix estimator that captures the variance improvement term. Fifth, we provide a functional central limit theorem analogous to \pc{koenker1999goodness} that holds uniformly with respect to the bandwidth. Sixth, we show that our smoothed QR slope estimator is asymptotically differentiable. We exploit this feature to come up with conditional quantile density function (qdf) and probability density function (pdf) estimators that do not suffer from the curse of dimensionality under correct QR specification. In particular, this allows us to propose an efficient QR estimator using sample splitting weights based on the qdf estimator.

Our simulations examine confidence intervals using a random bandwidth. Although it is as usual hard to estimate it due to the derivative of the conditional pdf that appears in the bias term, we show how to circumvent that by resorting to \pc{silverman1986density} rule-of-thumb bandwidth for the standard QR residuals. The simulation experiment also illustrates the estimation of the qdf and pdf in presence of a reasonably large number of explanatory variables.

We are obviously not the first to sail in the smoothing direction. In particular, \cn{nadaraya1964some} and \cn{parzen1979nonparametric} estimate unconditional quantiles respectively by inverting a smoothed estimator of the cumulative distribution function (cdf) and by smoothing the sample quantile function. \cn{xiang1994bahadur} and \cn{ralescu1997bahadur} provide Bahadur-Kiefer representations for smoothed quantile estimators. \cn{azzalini1981note} shows that the former smoothed estimator dominates sample quantiles at the second order \ca{cheung2010bootstrap}{see also}, with \cn{sheather1990kernel} establishing a similar result for the latter. For discussions on the relative deficiency of sample quantiles with respect to kernel-based quantile estimators, see also \cn{falk1984relative} and \cn{kozek2005combine}. Although the above quantile estimators are all continuous, only a few of them consider covariates in a quantile regression fashion. \cn{volgushev2019distributed} allow for covariates, ensuring quantile-level differentiability using spline smoothing.

The asymptotic covariance matrix of quantile estimators involves an unknown pdf \cite{koenker2005quantile}, making inference a bit harder.\footnote{~~For instance, \cn{hall1988distribution} study higher-order accuracy of confidence intervals. See also \cn{portnoy2012nearly} and references therein.} Smoothing the objective function helps in this dimension for it allows estimating the asymptotic covariance matrix using a standard sandwich formula \cite{kaplan2017smoothed}. In turn, we focus more on the implications of smoothing the objective function, notably in terms of smoothness of the QR estimator, and on quantile level and bandwidth uniformity.

There is notwithstanding an intimate connection between the above papers on inference and ours. Although the smoothed objective function of \cn{horowitz1998bootstrap} differs from ours, his theoretical arguments easily extends to our framework. In addition, the first-order conditions that \cn{whang2006smoothed} employs to derive his empirical likelihood QR estimator is identical to the ones our smoothed QR estimator solves. Interestingly, the same applies to \pc{kaplan2017smoothed} smoothed estimating equations (SEE) if restricting attention to QR without instrumental variables. As us, they argue that their smoothing approach can improve on \pc{horowitz1998bootstrap} smoothed QR estimator. See also \cn{kaido2018decentralization} for numerical algorithms that connect standard SEE and QR estimations.

Finally, our paper also relates with the literature on qdf estimation. \cn{parzen1979nonparametric} is the first to notice the important role that the quantile density plays. See \cn{guerre2012uniform} for a recent application in the econometrics of auctions, as well as \cn{roth2018estimating} and references therein for related estimation methods.\footnote{~~In particular, \cn{roth2018estimating} develop a qdf estimator based on the estimation of the second derivative of the population QR objective function.} We propose a novel efficient QR estimator based on a first-step qdf estimation, whose second-order behavior avoids the curse of dimensionality, as opposed to the estimation procedures of \cn{Newey1990efficient}, \cn{Zhao2001asymptotically}, \cn{otsu2008conditional}, and \cn{komunjer2010efficient}.

The remainder of this paper proceeds as follows. Section 2 introduces our smoothed QR estimator. Section 3 describes the main assumptions, the estimation procedure, the asymptotic covariance matrix estimator and main results. Section 4 assesses by means of a simulation study the performance of our kernel-based QR estimator relative to \pc{koenker1978regression} and \pc{horowitz1998bootstrap} estimators. Section 5 offers some concluding remarks. Appendix A collects the proofs of our main results, whereas we relegate the proofs of the intermediary results to an online appendix. The latter also includes an in-depth comparison with \pc{horowitz1998bootstrap} smoothing approach.

\section{Smoothed QR estimation and extensions}

Let $(Y_i,X_i)$, with $i=1,\ldots,n$, denote an iid sample from $(Y,X)\in\mathbb{R}\times\mathbb{R}^d$, where the conditional quantile of the response $Y$ given the covariate $X=x$ is such that
\begin{equation}\label{eq:linear-qr-model}
Q(\tau\given x)=x'\beta(\tau),\quad\tau\in\left(0,1\right),
\end{equation}
with $Q(\tau\given x):=\inf\big\{q:\,F(q\given x)\ge\tau\big\}$ and $F(\cdot\given x)$ denoting the conditional cdf of $Y$ given $X=x$, with density $f(\cdot\given x)$. We henceforth assume the correct specification of the QR model in \eqref{eq:linear-qr-model}.

\cn{koenker1978regression} define the (population) objective function of the quantile regression as
\begin{equation}\label{eq:obj-function-R}
R(b;\tau):=\E\left[\rho_\tau\big(e(b)\big)\right]=\int\rho_\tau(t)\,\mathrm{d}F(t;b),
\end{equation}
where $e(b):=Y-X'b$, $F\left(t;\,b\right)=\Prob\big[e(b)\le t\big]$, and $\rho_\tau(u):=u\big[\tau-\mathbb{I}(u<0)\big]$ is the usual check function with $\mathbb{I}(A)$ denoting the indicator function that takes value 1 if $A$ is true, zero otherwise. As the true parameter $\beta(\tau)$ minimizes \eqref{eq:obj-function-R}, \pc{koenker1978regression} standard QR estimator $\widehat{\beta}(\tau)$ minimizes the sample analog based on the empirical distribution, namely,
\begin{equation}\label{eq:obj-function-hat-R}
\widehat{R}(b;\tau):=\frac{1}{n}\sum_{i=1}^n\rho_\tau\big(e_i(b)\big)=\int\rho_\tau(t)\,\mathrm{d}\widehat{F}(t;b),
\end{equation}
where $e_i(b):=Y_i-X_i'b$, and $\widehat{F}(\cdot;b)$ denotes the empirical distribution function of $e_i(b)$.

The right-hand side of equation~\eqref{eq:obj-function-R} suggests that one may obtain a different QR estimator by varying the integrating measure of the check function integral, that is to say, by changing the estimator of the cdf. Instead of employing the empirical distribution as in the standard QR estimator, we shall consider a kernel-type cdf estimator as the integrating measure in a similar fashion to what \cn{nadaraya1964some} proposes for smoothing the unconditional quantile estimator.

Consider a bandwidth $h>0$ that shrinks to zero as the sample size grows and a smooth kernel function $k$ such that $\int k(v)\,\mathrm{d}v=1$. Letting $k_h(v)=\frac{1}{h}\,k(v/h)$, the kernel density and distribution estimators are given by $\widehat{f}_h(v;b):=\frac{1}{n}\sum_{i=1}^nk_h\big(v-e_i(b)\big)$ and $\widehat{F}_h(t;b):=\int_{-\infty}^t\widehat{f}_h(v;b)\,\mathrm{d}v$, respectively. We then apply kernel smoothing to the empirical objective function in \eqref{eq:obj-function-hat-R}, yielding
\begin{equation}\label{eq:obj-function-hat-R-h}
\widehat{R}_h(b;\tau):=\int\rho_\tau(t)\,\mathrm{d}\widehat{F}_h(t;b)=\int\rho_\tau(t)\,\widehat{f}_h(t;b)\,\mathrm{d}t.
\end{equation}
Accordingly, the resulting smoothed QR estimator is
\begin{equation}\label{eq:beta-argmin-smooth}
\widehat{\beta}_h(\tau):=\arg\min_{b\in\mathbb{R}^d}\widehat{R}_h(b;\tau).
\end{equation}

By smoothing the integrating measure of the objective function, we ensure that the mapping $b\mapsto\widehat{R}_h(b;\tau)$ is twice continuously differentiable, in opposition to the nonsmoothness of the standard objective function $b\mapsto\widehat{R}(b;\tau)$. The differentiability of the objective function is very convenient for at least two reasons. First, the smoothness of the objective function entails the regularity of the resulting QR estimator. Second, differentiability allows us to estimate the asymptotic covariance matrix of our quantile slope coefficient estimates in a canonical manner \ca{newey1994large}{see, for instance,}.

Finally, observe that
\begin{displaymath}
\widehat{R}_h(b;\tau)=(1-\tau)\int_{-\infty}^0\widehat{F}_h(v;b)\,\mathrm{d}v+\tau\int_0^\infty\big(1-\widehat{F}_h(v;b)\big)\,\mathrm{d}v,
\end{displaymath}
so that the first- and second-order derivatives of $\widehat{R}_h(b;\tau)$ with respect to $b$ are respectively
\begin{displaymath}
\widehat{R}_h^{(1)}(b;\tau)=\frac{1}{n}\sum_{i=1}^n X_i\left[K\left(-\frac{e_i(b)}{h}\right)-\tau\right]~~\textrm{and}~~ \widehat{R}_h^{(2)}(b;\tau)=\frac{1}{n}\sum_{i=1}^n X_iX_i'k_h\big(-e_i(b)\big),
\end{displaymath}
where $K(t):=\int_{-\infty}^tk(v)\,\mathrm{d}v$.

\subsection{Asymptotic inference}

The availability of the second-order derivative $\widehat{R}_h^{(2)}\big(\widehat{\beta}_h(\tau);\tau\big)$ allows us to propose asymptotic confidence intervals for $\beta(\tau)$ in a standard fashion. We show in Section~3.4 that $\sqrt{n}\big(\widehat{\beta}_h(\tau)-\beta(\tau)\big)$ weakly converges to a Gaussian distribution with mean zero and covariance matrix $\Sigma(\tau):=D^{-1}(\tau)V(\tau)D^{-1}(\tau)$, where $V(\tau):=\tau(1-\tau)\E(XX')$ and
\begin{displaymath}
D(\tau):=R^{(2)}\big(\beta(\tau);\tau\big)=\E\!\left[XX'f\left(X'\beta(\tau)\given X\right)\right]
\end{displaymath}
is the Hessian of the objective function evaluated at the true parameter. In addition, we show that
\begin{displaymath}
\widehat{\Sigma}_h(\tau):=\widehat{D}_h^{-1}(\tau)\widehat{V}_h(\tau)\widehat{D}_h^{-1}(\tau),
\end{displaymath}
with $\widehat{D}_h(\tau):=\widehat{R}_h^{(2)}\big(\widehat{\beta}_h(\tau);\tau\big)$ and
\begin{displaymath}
\widehat{V}_h(\tau):=\frac{1}{n}\sum_{i=1}^n X_iX_i'\left[K\left(-\frac{e_i\big(\widehat{\beta}_h(\tau)\big)}{h}\right)-\tau\right]^2,
\end{displaymath}
is a consistent estimator of $\Sigma(\tau)$.\footnote{~~Alternatively, one could simply employ the sample analog of $\tau(1-\tau)\E[XX']$ to estimate $V(\tau)$, which would lead to \pc{powell1991} estimator of the asymptotic covariance of the standard QR estimator. See also \cn{angrist2006} and \cn{kato2012}. However, we expect that our variance estimator to entail better finite-sample properties and, accordingly, shorter confidence intervals.} This means that we may compute a $(1-\alpha)-$confidence interval for the $k$th QR coefficient $\beta_k(\tau)$ as
\begin{displaymath}
\mathrm{CI}_{1-\alpha}\big(\beta_k(\tau)\big):=\widehat{\beta}_{k,h}(\tau)\pm\frac{z_{\alpha/2}\,\widehat{\sigma}_{k,h}(\tau)}{\sqrt{n}},
\end{displaymath}
where $\widehat{\sigma}_{k,h}(\tau)$ is the square root of the $k$th diagonal entry of $\widehat{\Sigma}_h(\tau)$, $\widehat{\beta}_{k,h}(\tau)$ is the $k$th element of the smoothed QR estimator, and $z_{\alpha}$ is the $\alpha$ quantile of a standard Gaussian distribution.

\subsection{Smoothness of $\widehat{\beta}_h (\cdot)$ and extensions\label{sec:pdf-estimator}}

Given that $\widehat{\beta}_h(\tau)$ satisfies the first-order condition $\widehat{R}_h^{(1)}\big(\widehat{\beta}_h(\tau);\tau\big)=0$, it follows from the implicit function theorem that $\widehat{\beta}_h(\tau)$ is continuously differentiable with respect to $\tau$ with
\begin{equation}
\frac{\partial\widehat{\beta}_h(\tau)}{\partial\tau}=-\left[\widehat{R}_h^{(2)}\big(\widehat{\beta}_h(\tau);\tau\big)\right]^{-1} \frac{\partial\widehat{R}_h^{(1)}\big(\widehat{\beta}_h(\tau);\tau\big)}{\partial\tau}=\left[\widehat{R}_h^{(2)} \big(\widehat{\beta}_h(\tau);\tau\big)\right]^{-1}\bar{X}=:\widehat{\beta}_h^{(1)}(\tau)
\label{Dbeta}
\end{equation}
as long as the Hessian has an inverse, which holds with a probability tending to $1$. Because the Hessian matrix is asymptotically positive, it also follows that the QR function evaluated at the average covariate, $\bar{X}'\widehat{\beta}_h(\tau)$, is strictly increasing in $\tau$.

Figure 1 compares two different paths of our smoothed QR estimator with the ones of the standard QR estimator. The quantile paths implied by the convolution-type kernel QR estimates are not only much smoother but much less variable than the conditional quantiles based on the standard QR estimates. Theorem 2 indeed shows that our smoothed QR estimator is continuous with a probability tending to one, while the standard QR estimator is a step function. Theorem 3 also establishes that smoothing reduces the variance of the QR estimator. Even though Figure 1 evaluates the conditional quantiles at $X=0.1$ and $X=0.9$, rather than at $\bar{X}$, the quantile paths based on our smoothed QR estimator look increasing such that there is no need to apply \pc{CFVG10} rearrangement procedure. This is most likely due to the smoothness of the convolution-type kernel estimator, inasmuch as the nonmonotonicity of the standard QR estimator is mostly due to small jumps.\vfill

We next consider two extensions based on the quantile density function (qdf) given by
\begin{displaymath}
q(\tau|x)=\frac{\partial Q(\tau|x)}{\partial\tau}=\frac{1}{f\big(Q(\tau|x)|x\big)}.
\end{displaymath}
As noted by \cn{GS2017}, the curve $\tau\mapsto\big(Q(\tau|x),1/q(\tau|x)\big)=:\mathtt{f}(\tau|x)$ is the graph $y\mapsto\big(y,f(y|x)\big)$ of the conditional pdf $f(\cdot|x)$. The qdf plays a major role in first-price auctions \cite{guerre2012uniform} and in semiparametric efficient QR estimation \ca{Newey1990efficient,Zhao2001asymptotically}{see, among others,}. In particular, the efficient QR slope estimator $\tilde{b}_q(\tau)=\arg\min_b\widetilde{R}_q(b;\tau)$ is infeasible because
\begin{equation}\label{Infefficient}
\widetilde{R}_q(b;\tau)=\frac{1}{n}\sum_{i=1}^n\frac{\rho_{\tau}(Y_i-X_i'b)}{q(\tau|X_i)}
\end{equation}
depends on the conditional qdf. \cn{Newey1990efficient} and \cn{Zhao2001asymptotically} propose to estimate the conditional qdf using kernel methods. However, their estimators are bound to perform poorly if the covariate dimension is large due to the curse of dimensionality. The same drawback also applies to the efficient estimators put forth by \cn{otsu2008conditional} and \cn{komunjer2010efficient}.

\paragraph{Pdf curve estimation.}~~~~The linear QR model dictates that $q(\tau|x)=x'\beta^{(1)}(\tau)$, which we may estimate using $\widehat{\beta}_h^{(1)}(\tau)=\left[\widehat{R}_h^{(2)}\big(\widehat{\beta}_h(\tau);\tau\big)\right]^{-1}\bar{X}$ in (\ref{Dbeta}). The resulting conditional qdf estimator, $\widehat{q}_h(\tau|x):=x'\widehat{\beta}_h^{(1)}(\tau)$, does not suffer from the curse of dimensionality because it makes use of the linear QR structure. The pdf curve estimator
\begin{equation}\label{eq:density-estimator}
\widehat{\mathtt{f}}_h(\tau|x):=\left(x'\widehat{\beta}_h(\tau),\frac{1}{x'\widehat{\beta}_h^{(1)}(\tau)}\right)
\end{equation}
converges to $\mathtt{f}(\tau|x)$ at the rate of univariate kernel density estimator. If $\widehat{q}_h(\cdot|x)>0$, the pdf estimator is positive and integrates to one given that $\int_0^11/\widehat{q}_h(\tau|x)\,\mathrm{d}\widehat{Q}_h(\tau|x)=\int_0^1\mathrm{d}\tau=1$. This means that the pdf curve estimator is strict. Figure \ref{fig:pdfonedraw} depicts an example of pdf curve estimation with three covariates using a sample of 200 observations. It also exhibits the QR and qdf estimates we use for computing the pdf estimator. The pdf curve estimator seems to capture reasonably well the underlying density despite the high dimensionality of the estimation problem relative to the sample size.

\paragraph{Semiparametric efficiency.}~~~~As in \cn{Newey1990efficient} and \cn{Zhao2001asymptotically}, we resort to sample splitting to obtain efficient QR slope estimators. To this end, we minimize a feasible counterpart of the objective function $\widetilde{R}_q(b;\tau)$ in \eqref{Infefficient}, namely,
\begin{equation}
\check{R}_q(b;\tau)=\frac{1}{n-m}\sum_{i=m+1}^n\frac{\rho_{\tau}\left(Y_i-X_i'b\right)}{\check{q}_h(\tau|X_i)},\label{Weighted}
\end{equation}
where $\check{q}_h(\tau|x)=x'\check{\beta}_h^{(1)}(\tau)$ considers only the first $m<n$ observations. One could alternatively use either a smoothed version of the check function $\rho_{\tau}(\cdot)$ or a version of \pc{Newey1990efficient} one-step estimator using the weights $\check{q}_h(\tau|X_i)$.

Theoretical results for pdf graph and for the efficient QR in (\ref{Weighted}) readily follow from the asymptotic theory we establish for the smoothed QR estimator. We state them in Section \ref{S:extensions}.

\subsection{Alternative smoothed objective function}

In the absence of covariates, the first-order condition $\widehat{R}_h^{(1)}\big(\widehat{\beta}_h(\tau);\tau\big)=0$ that results in our smoothed QR estimator coincides with the first-order condition in \cn{nadaraya1964some}, namely, $\widehat{F}_h\big(\widehat{\beta}_h(\tau)\big)=\tau$. This confirms that the smoothing we apply to the objective function boils down to estimating the conditional cumulative distribution function using a kernel approach. In turn, the smoothed objective function $\widehat{\mathfrak{R}}_h(b;\tau)$ put forth by \cn{horowitz1998bootstrap} replaces the indicator in the check function by a kernel counterpart: $\widehat{\mathfrak{R}}_h(b;\tau):=\frac{1}{n}\sum_{i=1}^ne_i(b)\Big[\tau-K\big(-e_i(b)/h\big)\Big]$. As noted by \cn{kaplan2017smoothed}, the first-order derivative of this objective function is
\begin{eqnarray}
\widehat{\mathfrak{R}}^{(1)}_h(b;\tau)&=&\frac{1}{n}\sum_{i=1}^nX_i\left[K\left(-\frac{e_i(b)}{h}\right)-\tau-k\left(-\frac{e_i(b)}{h}\right)\frac{e_i(b)}{h}\right]\notag\\
                                      &=&\widehat{R}^{(1)}_h(b;\tau)-\frac{1}{n}\sum_{i=1}^nX_ik\left(-\frac{e_i(b)}{h}\right)\frac{e_i(b)}{h}.\label{horowitz.obj.func}
\end{eqnarray}
The resulting first-order condition differs from the one we obtain because of the additional term in \eqref{horowitz.obj.func}. We show in the Online Appendix that, as a consequence, \pc{horowitz1998bootstrap} smoothed QR estimator exhibits larger bias and variance than ours.

\section{Asymptotic theory}

We start with some notation. Let $\norm{\cdot} $ denote the Euclidean norm of a matrix, namely, $\norm{A}=\sqrt{\trace(AA')}$. We denote by $f(\cdot\given x)$ the conditional probability density function of $Y$ given $X=x$, with $j$th partial derivative given by $f^{(j)}(y\given x):=\frac{\partial^j}{\partial y^j}\,f(y\given x)$. Similarly, let $q(\tau\given x):=\frac{\partial}{\partial\tau}\,Q(\tau\given x)$. In what follows, we first discuss the assumptions we require to work out the asymptotic theory and then derive the asymptotic mean squared error and Bahadur-Kiefer representation for our smoothed QR estimator. We wrap up this session with some inference implications.

\subsection{Assumptions}

In this section, we discuss the conditions under which we derive the asymptotic theory. Apart from standard technical conditions on the covariates and kernel function, we essentially require the conditional quantile and density functions to be smooth enough. In particular, we assume the following conditions.\vskip 1em

\noindent\textbf{Assumption~X}~~~~The components of $X$ are positive, bounded random variables, i.e. the support of $X$ is a compact subset of $\bar{\mathbb{R}}_{+*}^d$. The matrix $\E[XX']$ is full rank.\vskip 1ex

\noindent\textbf{Assumption~Q}~~~~The conditional quantile and density functions $Q(\tau\given x)$ and $f(y\given x)$ satisfy
\begin{description}
\item[Q1~] The map $\tau\mapsto\beta(\tau)$ is continuously differentiable over $(0,1)$. The conditional density $f(y\given x)$ is continuous and strictly positive over $\mathbb{R}\times\supp(X)$.
\item[Q2~] There exists an integer $s\ge 1$ such that the derivative $f^{(s)}(\cdot\given\cdot)$ is uniformly continuous in the sense that $\lim_{\epsilon\rightarrow 0}\sup_{(x,y)\in\mathbb{R}^{d+1}} \sup_{t:\,\abs{t}\le\epsilon}\abs{f^{(s)}(y+t\given x)-f^{(s)}(y\given x)}=0$, as well as such that $\sup_{(x,y)\in\mathbb{R}^{d+1}}\abs{f^{(j)}(y\given x)}<\infty$ and $\lim_{y\rightarrow\pm\infty}f^{(j)}(y\given x)=0$ for all $j=0,\ldots,s$.
\end{description}\vskip 1ex

\noindent\textbf{Assumption~K}~~~~The kernel function $k$ and bandwidth $h$ satisfy
\begin{description}
\item[K1~] The kernel $k:\mathbb{R}\rightarrow\mathbb{R}$ is even, integrable, twice differentiable with bounded first and second derivatives, and such that $\int k(z)\,\mathrm{d}z=1$ and $0<\int_0^\infty K(z)\,[1-K(z)]\,\mathrm{d}z<\infty$. In addition, for $s$ as in Assumption~Q2, $\int\abs{z^{s+1}k(z)}\,\mathrm{d}z<\infty$, and $k$ is orthogonal to all nonconstant monomials of degree up to $s$, i.e., $\int z^jk(z)\,\mathrm{d}z=0$ for $j=1,\dots,s$, and $\int z^{s+1} k(z)\,\mathrm{d}z \neq 0$.
\item[K2~] $h\in[\ubar{h}_n,\bar{h}_n]$ with $1/\ubar{h}_n=o\big((n/\ln n)^{1/3}\big)$ and $\bar{h}_n=o(1)$.
\end{description}\vskip 1em

Some remarks are in order. First, observe that $R^{(2)}(b;\tau)=\E\!\left[XX'f\!\left(X'b\given X\right)\right]$ is positive definite for all $b$ and any $\tau$ under Assumptions~Q1 and X. This means that $D^{-1}(\tau)$ exists for every $\tau$. Second, Assumption~Q1 also ensures that $\tau\mapsto Q(\tau\given x)$ is strictly increasing over $\left(0,1\right)$, with a strictly positive derivative with respect to $\tau$ given that $q(\tau\given x)=1/f\big(Q(\tau\given x)\given x\big)$. Third, we assume that $Y$ has support on the real line merely for notational simplicity. It is straightforward to relax it with some minor adaptations. Fourth, the reason why we choose a kernel such that $\int_0^\infty K(z)\,[1-K(z)]\,\mathrm{d}z$ is positive will become clear in Theorem~3. It makes sense because it guarantees that our smoothed QR estimator dominates the standard QR estimator in the AMSE sense. Fifth, in view that Assumption~K does not preclude high-order kernels, $\widehat{f}_h\left(\cdot;b\right)$ is not necessarily a density, even if it is a consistent estimator of the density of the error term $e(b)$. It is nonetheless possible to show $\widehat{R}_h(b;\tau)=\frac{1}{n}\sum_{i=1}^n\rho_\tau*k_h\big(e_i(b)\big)$, where $*$ is the convolution operation. This means that, technically speaking, it is probably more rigorous to interpret our approach as a convolution-type smoothing. Finally, it is also important to clarify that the bandwidth $h$ implicitly depends on the sample size $n$ through its lower and upper limits in Assumption~K2. This is paramount because we wish to entertain data-driven bandwidths.

\subsection{Bias and Bahadur-Kiefer representation}

In this section, we study the order of the asymptotic bias and obtain a Bahadur-Kiefer representation for the stochastic error of $\widehat{\beta}_h(\tau)$. As for the former, it is indeed expected from popular wisdom that smoothing the QR objective function should induce bias in finite samples. More formally, $\widehat{\beta}_h(\tau)$ actually estimates $\beta_h(\tau):=\arg\min_{b\in\mathbb{R}^d}R_h(b;\tau)$, with $R_h(b;\tau):=\E[\widehat{R}_h (b;\tau)]$. This yields $\beta_h(\tau)-\beta(\tau)$ as the bias term and $\widehat{\beta}_h(\tau)-\beta_h(\tau)$ as the stochastic error of our smoothed QR estimator. Our first result shows that the smoothing bias shrinks to zero as the sample size grows.\vskip 1em

\noindent\textbf{Theorem 1}~~~~\emph{For $\bar{h}_n$ small enough, $\beta_h(\tau)$ is unique under Assumptions~X, Q and K for every $\tau\in[\ubar{\tau},\bar{\tau}]$ and such that $\beta_h(\tau)=\beta(\tau)-h^{s+1}B(\tau)+o(h^{s+1})$ uniformly over $(\tau,h)\in[\ubar{\tau},\bar{\tau}]\times[\ubar{h}_n,\bar{h}_n]$, with $B(\tau)=\frac{\int z^{s+1}k(z)\,\mathrm{d}z}{(s+1)!}\, D^{-1}(\tau)\,\E\!\left[Xf^{(s)}\big(X'\beta(\tau)\given X\big)\right]$.}\vskip 1em

It is interesting to discuss the implications of Theorem 1 to the particular case of a standard linear regression model $Y_i=X_i' \beta +\varepsilon_i$ with iid errors independent of the covariates. Let $X_i=(1,\tilde{X}_i)'$, with $\tilde{X}_i\in\mathbb{R}$, so that the conditional pdf of $Y_i$ given $X_i=x$ is $f(y|x)=f_\varepsilon\big(y-x'\beta(\tau)\big)$. It then follows that  $D(\tau)=f_\varepsilon(0)\E[XX']$ for all $\tau$ and $B(\tau)\propto\E^{-1}[XX']\E[X]=(1,0)'$, so that the first term in the bias appears only for the intercept. More generally, Theorem~1 settles the issue of possible side effects of smoothing in that $\beta_h(\tau)$ eventually becomes uniformly close to the true parameter $\beta(\tau)$.

The next result derives some convenient expansions for the stochastic error $\widehat{\beta}_h(\tau)-\beta_h(\tau)$. For this purpose, let $\widehat{S}_h(\tau):=\widehat{R}_h^{(1)}\big(\beta_h(\tau);\tau\big)$ and $D_h(\tau):=R_h^{(2)}\big(\beta_h(\tau);\tau\big)$. Note that the first-order condition $R_h^{(1)}\big(\beta_h(\tau);\tau\big)=0$ implies that the score term $\widehat{S}_h(\tau)$ has zero mean, and hence the stochastic error in the Bahadur-Kiefer representation (\ref{eq:bahadur-representation}) is asymptotically centered.\vskip 1em

\noindent\textbf{Theorem 2}~~~~\emph{Under Assumptions~X, Q and K, $\widehat{\beta}_h(\cdot)$ is unique and continuous over $(\tau,h)\in[\ubar{\tau},\bar{\tau}]\times[\ubar{h}_n,\bar{h}_n]$ with probability tending to one, satisfying the following two representations:
\begin{eqnarray}
\sqrt{n}\,\left(\widehat{\beta}_h(\tau)-\beta_h(\tau)\right)&=&-\sqrt{n}D_h^{-1}(\tau)\,\widehat{S}_h(\tau)+O_p\big(\varrho_n(h)\big)\label{eq:bahadur-representation}\\
                                                            &=&-\sqrt{n}\widehat{D}_h^{-1}(\tau)\widehat{S}_h(\tau)+O_p(1/\sqrt{n})\label{eq:nonlinear-representation}
\end{eqnarray}
with $\varrho_n(h)=\sqrt{\ln n/(nh)}$  and both remainder terms uniform with respect to $(\tau,h)\in[\ubar{\tau},\bar{\tau}]\times[\ubar{h}_n,\bar{h}_n]$.}\vskip 1em

 Unlike the standard QR objective function, $\widehat{R}_h (\cdot;\tau)$ is not necessarily convex because higher-order kernels may take negative values. We address this issue by first showing that $\widehat{\beta}_h(\tau)$ is close to $\beta_h(\tau)$, uniformly in $\tau$ and $h$, and then proving that the smoothed objective function $\widehat{R}_h (\cdot;\tau)$ is asymptotically strictly convex in the vicinity of $\beta_h(\tau)$. This follows from the convergence of $\widehat{R}^{(2)}_h(b;\tau)$ to $R^{(2)}(b;\tau)$, uniformly for $b$ in any compact set and also in $\tau$ and $h$, as established using a powerful concentration inequality from \cn{massart2007concentration}. The first-order condition $\widehat{R}_h\big(\widehat{\beta}_h(\tau);\tau\big)=0$ implies the following integral representation
\begin{displaymath}
	\widehat{\beta}_h(\tau)-\beta_h(\tau)=-\left[\int_0^1\widehat{R}_h^{(2)}\left(\beta_h(\tau)+u\big[\widehat{\beta}_h(\tau)-\beta_h(\tau)\big];\tau\right)\,\mathrm{d}u\right]^{-1}\widehat{S}_h^{(1)}\big(\tau\big)
\end{displaymath}
for the stochastic error. Deriving the order of the score uniformly in $\tau$ and $h$ via a concentration inequality entails the representations (\ref{eq:bahadur-representation}) and (\ref{eq:nonlinear-representation}). One could also obtain higher-order expansions for the stochastic error along the same lines by establishing uniform convergence of higher-order derivatives of the smooth objective function through concentration inequalities.

The Bahadur-Kiefer representation in \eqref{eq:bahadur-representation} shows that $\widehat{\beta}_h(\tau)$ is, in a sense, more linear than the standard QR estimator $\widehat{\beta}(\tau)$. \cn{knight2001comparing} and \cn{jurevckova2012methodology} show that, in many cases of interest, the Bahadur-Kiefer representation for the standard QR estimator is given by $\sqrt{n}\left(\widehat{\beta}(\tau)-\beta(\tau)\right)=-\sqrt{n}\,D^{-1}(\tau)\widehat{S}(\tau)+O_p(n^{-1/4})$, with $\widehat{S}(\tau)=\widehat{R}^{(1)}\big(\beta(\tau);\tau\big)$. In contrast, the remainder term in \eqref{eq:bahadur-representation} is of order nearly $O_p(n^{-1/2})$ for proper bandwidth choices if centering around $\beta_h(\tau)$. The asymptotically negligible bias given by the difference between $\beta_h(\tau)$ and $\beta(\tau)$ is the price we pay for improving the rate of the Bahadur-Kiefer representation.

The nonlinear approximation in \eqref{eq:nonlinear-representation} obtains a smaller remainder term of order $n^{-1/2}$ by replacing the deterministic standardization $D_h^{-1}(\tau)$ in \eqref{eq:bahadur-representation} with $\widehat{D}_h^{-1}(\tau)$. This slightly improves the rate of the remainder term in \pc{portnoy2012nearly} approximation. One could also obtain higher-order approximations involving quadratic terms under stronger bandwidth rate conditions. As in \cn{horowitz1998bootstrap}, this would however exclude bandwidths of the optimal order $n^{-1/5}$ for a kernel of order $s+1=2$, and hence we do not follow such alternative. Lemma~4 in the Appendix shows that \eqref{eq:nonlinear-representation} also holds for $\widehat{R}^{(2)}\big(\beta_h(\tau);\tau\big)$ in lieu of $\widehat{D}_h(\tau)$. Using the standardization $\widehat{D}_h(\tau)$ instead of the infeasible
$\widehat{R}^{(2)}\big(\beta_h(\tau);\tau\big)$	is important for practical purposes, such as computing Wald statistics.

It also follows from Theorems 1 and 2 that $\widehat{\beta}_h(\cdot)$ offers a fair global picture of $\beta(\cdot)$ in that
\begin{equation}\label{eq:consistency-uniform}
\norm{\widehat{\beta}_h(\tau)-\beta(\tau)}=O_p\left(\frac{1}{\sqrt{n}}+{h}^{s+1}\right),
\end{equation}
uniformly for $\tau\in[\ubar{\tau},\bar{\tau}]$ and $h\in[\ubar{h}_n,\bar{h}_n]$. Accordingly, the remainder in \eqref{eq:consistency-uniform} is of order $O_p(n^{-1/2})$ for any $h\le O\left(n^{-1/(2(s+1))}\right)$. The latter restriction is quite light, holding not only for the infeasible AMSE-optimal bandwidth in Theorem~4 but also for the corresponding rule-of-thumb bandwidth that we suggest in Section \ref{S:AMSE}.

Lastly, it is important to stress the major role that uniformity plays here. It ensures that, if a random (possibly data-driven) bandwidth process $\left\{\widehat{h}(\tau);\,\tau\in[\ubar{\tau},\bar{\tau}]\right\}$ has sample paths in $[\ubar{h}_n,\bar{h}_n]$ with a sufficiently high probability, then \eqref{eq:bahadur-representation} remains valid even if we replace $h$ with a data-driven bandwidth $\widehat{h}(\tau)$. The next result states this property in a rigorous manner.\vskip 1em

\noindent\textbf{Corollary 1}~~~~\emph{If $\left\{\widehat{h}(\tau);\,\tau\in[\ubar{\tau},\bar{\tau}]\right\}$ satisfies $\Prob\!\left(\widehat{h}(\tau)\in[\ubar{h}_n,\bar{h}_n]~\mathrm{for~all}~\tau\right)\rightarrow 1$, then both \eqref{eq:bahadur-representation} and \eqref{eq:nonlinear-representation} hold with $\widehat{h}(\tau)$ in place of $h$, uniformly in $\tau$.}\vskip 1em

The asymptotic theory so far posits that our smoothed QR estimator entails a better Bahadur-Kiefer representation than the standard QR estimator and that we may employ a data-driven bandwidth that depends on the quantile level and covariates. In the next section, we complement the asymptotic theory by characterizing the AMSE of our convolution-type kernel QR estimator as well as the bandwidth choice that minimizes it.

\subsection{Asymptotic mean squared error}\label{S:AMSE}

The asymptotic covariance matrix of $\widehat{\beta}_h(\tau)$ comes from the leading term of its Bahadur-Kiefer linear representation in \eqref{eq:bahadur-representation}. The next result not only characterizes this asymptotic covariance matrix but also shows that it is smaller than the asymptotic covariance matrix of $\widehat{\beta}(\tau)$. In what follows, let $\Sigma_h(\tau):=\V\!\left(\sqrt{n}\,D_h^{-1}(\tau)\widehat{S}_h(\tau)\right)$.\vskip 1em

\noindent\textbf{Theorem 3}~~~~\emph{Assumptions~X, Q and K ensure that
\begin{equation}\label{eq:asymptotic-variance}
\Sigma_h(\tau)=\Sigma(\tau)-c_k\,h\,D^{-1}(\tau)+o(h),
\end{equation}
with $c_k=2\int_0^\infty K(y)[1-K(y)]\,\mathrm{d}y>0$, uniformly with respect to $(\tau,h)\in[\ubar{\tau},\bar{\tau}]\times[\ubar{h}_n,\bar{h}_n]$.}\vskip 1em

Theorem 3 shows that the asymptotic covariance matrix of the smoothed QR estimator is equal to the asymptotic covariance matrix of the standard QR estimator minus a term $c_k\,h\,D^{-1}(\tau)$ induced by smoothing. Table~\ref{tab:kernel} reports the values of $c_k$ for Gaussian-type kernels that satisfy Assumption~ K1. The order $\ln n/(nh)$ of the squared Bahadur-Kiefer remainder term is negligible with respect of the order $h$ of the smoothing-related term given that $1/h=o(\sqrt{n/\ln n})$ by Assumption~K2.\vskip 1em

It is sometimes possible to refine the asymptotic covariance matrix expansion in Theorem 3 using strong approximation tools. For instance, assume there is no covariate and that $Y_i$ is uniform over $[0,1]$. In this case, there is no bias in that $\beta_h(\tau)=\tau$ and $D_h(\tau)=D(\tau)=1$ for $h$ small enough. The score function then reads
\begin{displaymath}
\widehat{S}_h(\tau)=\int\frac{1}{\sqrt{n}}\sum_{i=1}^n\Big[\mathbb{I}(Y_i\le\tau+ht)-(\tau+ht)\Big]k(t)\,\mathrm{d}t,
\end{displaymath}
using $\int k(t)\,\mathrm{d}t=1$ and $\int tk(t)\,\mathrm{d}t=0$. The Koml\'os-Major-Tusn\'ady strong approximation result shows that one may reconstruct the sample jointly with a sequence of Brownian bridges $B_n(\cdot)$ such that $\sup_{y\in[0,1]}\left|\frac{1}{\sqrt{n}}\sum_{i=1}^n\Big[\mathbb{I}(Y_i\le\tau+ht)-(\tau+ht)\Big]-B_n(t)\right|=O_p\!\left(\frac{\ln n }{\sqrt{n}}\right)$; see \cn{pollard2002user}. It follows that, if $k$ has a compact support,
\begin{eqnarray}
\widehat{S}_h(\tau)&=&\int B_n(\tau+ht)k(t)\,\mathrm{d}t+O_p\left(\frac{\ln n}{\sqrt{n}}\right)\nonumber\\
                   &=&B_n(\tau)+\int\Big[B_n(\tau+ht)-B_n(\tau)\Big]k(t)\,\mathrm{d}t+O_p\left(\frac{\ln n }{\sqrt{n}}\right). \label{eq:KMT}
\end{eqnarray}
As the integral term that comes from smoothing vanishes for $h=0$, $B_n(\tau)$ gives the limit distribution of the standard QR estimator.

The covariance between $B_n(\tau)$ and $B_n(\tau+ht)-B_n(\tau)$ is negative of order $h$. This implies a negative covariance between $B_n(\tau)$ and the integral term in \eqref{eq:KMT}. As the variance of the latter is of smaller order than this negative covariance, the asymptotic variance of $\widehat{S}_h(\tau)$ is smaller than $\mathbb{V}(B_n(\tau))=\tau(1-\tau)$ by a term of order $h$. Theorem 3 also implies that the distribution of $D_h(\tau)^{-1}\widehat{S}_h(\tau)$ is a centered normal with variance $\tau(1-\tau)-c_k h$ up to the order $\ln n /\sqrt{n}$. Because the second term in \eqref{eq:asymptotic-variance} is negative, the asymptotic variance of $\widehat{\beta}_h(\tau)$ is indeed smaller than the asymptotic variance of $\widehat{\beta}(\tau)$.

We next focus on obtaining the bandwidth $h_\lambda^*$ that minimizes the asymptotic mean squared error of $\lambda'\widehat{\beta}_h(\tau)$ for a given $\lambda\in\mathbb{R}^d$: $\AMSE\big(\lambda'\widehat{\beta}_h(\tau)\big)=\E\!\left[\lambda'\big(\beta_h(\tau)-D_h^{-1}(\tau)\widehat{S}_h(\tau)-\beta(\tau)\big)\right]^2$. This approximates the mean squared error $\MSE\big(\lambda'\widehat{\beta}_h(\tau)\big)=\E\!\left[\lambda'\big(\widehat{\beta}_h(\tau)-\beta(\tau)\big)\right]^2$ by essentially ignoring the asymptotically negligible remainder term of the Bahadur-Kiefer representation. To do so, we require that the bandwidth is such that $1/h=o\!\left(\sqrt{n/\ln n}\right)$; see Theorems~2 and 3.\vskip 1em

\noindent\textbf{Theorem 4}~~~~\emph{Let Assumptions~X, Q and K hold. If $\lambda'B(\tau)\ne 0$, and the conditional density $f(\cdot\given x)$ is $s$-times continuously differentiable for all $x$, then $\AMSE\big(\lambda'\widehat{\beta}_h(\tau)\big)$ is minimal for
\begin{equation}
h_\lambda^*=\left(\frac{c_k\,\lambda'D^{-1}(\tau)\lambda}{2n(s+1)\big(\lambda'B(\tau)\big)^2}\right)^{\frac{1}{2s+1}},\label{eq:optimal.bandwidth}
\end{equation}
and equal to $\AMSE\big(\lambda'\widehat{\beta}_{h_\lambda^*}(\tau)\big)=\frac{1}{n}\,\lambda'\left[\Sigma(\tau)-c_k\,h_\lambda^*\,\frac{2s+1}{2s+2}\,D^{-1}(\tau)\right] \lambda+o(h_\lambda^*/n)$.}\vskip 1em

The optimal bandwidth we derive in Theorem~4 obviously depends on $\lambda$. This is not the case of \pc{kaplan2017smoothed} bandwidth choice, which minimizes $\E\Big[\big\Vert\widehat{\beta}(\tau)-\beta(\tau)\big\Vert^2\Big]$. Although their optimal bandwidth has the same order of $h_\lambda^*$ in \eqref{eq:optimal.bandwidth}, it obviously does not depend on $\lambda$ given that they focus on the global estimation of $\beta(\tau)$. As such, it is suboptimal for a linear combination of the elements of $\beta(\tau)$, as we consider here. This matters for instance in the particular case of a standard linear regression in view that the optimal bandwidth is well defined only for the intercept as the leading term of the bias vanishes for the slope coefficients.

It is easy to appreciate that Theorem 4 remains valid for any bandwidth $h_\lambda=[1+o(1)]h_\lambda^*$. Although entertaining a plug-in bandwidth based on the nonparametric estimation of $D(\tau)$ and $B(\tau)$ is certainly feasible, it is perhaps not very advisable due to the presence of the pdf derivative in the bias term. We circumvent this issue by means of a rule-of-thumb approach. In particular, it follows from the objective function in (\ref{eq:obj-function-hat-R-h}) that it is as if our convolution-kernel QR estimator were using a kernel-based estimator of the cdf of the standard QR residuals. This naturally leads to using \pc{silverman1986density} rule-of-thumb bandwidth based on residual dispersion measures (e.g., sample standard deviation or interquantile range).

\subsection{Inference}

In this section, we derive a functional limit distribution of the smoothed QR estimator that holds for data-driven bandwidths. We also discuss how to consistently estimate the asymptotic covariance matrix of $\widehat{\beta}_h(\tau)$ in order to make inference. In particular, we establish not only that $\widehat{\Sigma}_h(\tau):=\widehat{D}_h^{-1}(\tau)\widehat{V}_h(\tau)\widehat{D}_h^{-1}(\tau)$, with $\widehat{D}_h(\tau):=\widehat{R}_h^{(2)}\big(\widehat{\beta}_h(\tau);\tau\big)$ and
\begin{displaymath}
\widehat{V}_h(\tau):=\frac{1}{n}\sum_{i=1}^nX_iX_i'\left[K\left(-\frac{e_i\big(\widehat{\beta}_h(\tau)\big)}{h}\right)-\tau\right]^2,
\end{displaymath}
is a consistent estimator of $\Sigma(\tau)$, but also the asymptotic normality of our convolution-type kernel QR estimator.\vskip 1em

\noindent\textbf{Theorem 5}~~~~\emph{Let Assumptions~X, Q and K hold.}
\begin{description}
\item[(G)\,] \emph{$\Big\{\sqrt{n}\big(\widehat{\beta}_h(\tau)-\beta_h(\tau)\big):\,\tau\in[\ubar{\tau},\bar{\tau}]\Big\}$ converges in distribution to a centered Gaussian process $\big\{W(\tau):\,\tau\in[\ubar{\tau},\bar{\tau}]\big\}$ with covariance structure
	\begin{equation}\label{eq:asymptotic-gaussian-covariance}
    \V\big(W(\tau),W(\varsigma)\big)=(\tau\wedge\varsigma-\tau\varsigma)D(\tau)^{-1}\E(XX')D(\varsigma)^{-1},\qquad\tau,\varsigma\in[\ubar{\tau},\bar{\tau}].
	\end{equation}}
\item[(S)\,] \emph{Consider a data-driven bandwidth $\widehat{h}_n$ which satisfies $\widehat{h}_n=h_0\left\{1+o_p(1/\sqrt{h_n \ln n})\right\}$ for some deterministic $h_n$ in $[\ubar{h}_n,\overline{h}_n]$. $\Big\{\sqrt{n}\big(\widehat{\beta}_{\,\widehat{h}_n}(\tau)-\beta_{\,\widehat{h}_n}(\tau)\big):\,\tau\in[\ubar{\tau},\bar{\tau}]\Big\}$ converges in distribution to  $\big\{W(\tau):\,\tau\in[\ubar{\tau},\bar{\tau}]\big\}$.}
\item[(U)\,] \emph{If $\ubar{h}_n=o(1/\ln n)$, the convergence result in \textbf{(G)} is uniform in that
	\begin{displaymath}
	\sup_{h\in[\ubar{h}_n,\overline{h}_n]}\norm{\sqrt{n}\left(\widehat{\beta}_h(\tau)-\beta_h(\tau)-\left[\widehat{\beta}_{h_n}(\tau)-\beta_{h_n}(\tau)\right]\right)}=o_p(1)
	\end{displaymath}
    for any arbitrary deterministic sequence $h_n$ in $[\ubar{h}_n,\overline{h}_n]$.}
\item[(V)\,] \emph{$\widehat{\Sigma}_h(\tau)=\Sigma_h(\tau)+O_p\left(\sqrt{\ln n/(nh)}\right)+o(h^s)$ uniformly for $(\tau,h)\in[\ubar{\tau},\bar{\tau}]\times[\ubar{h}_n,\bar{h}_n]$.}		
\end{description}\vskip 1em

Theorem 5(G) shows that the functional limit distribution of the smoothed and standard QR estimators are identical.  Theorem 5(S) posits that such a functional central limit theorem (FCLT) holds for data-driven bandwidths that converge to a deterministic counterpart. Theorem 5(U) states that the FCLT holds uniformly with respect to the bandwidth. Theorem 5(V) implies not only that the confidence intervals of Section 2.1 asymptotically have the desired level, but also that $\widehat{\Sigma}_h(\tau)$ is a consistent estimator of the asymptotic covariance matrix of both standard and smoothed QR estimators. Under Assumption~K2, as $s\ge 1$,
\begin{displaymath}
\widehat{\Sigma}_h(\tau)=\Sigma_h(\tau)+o_p(h)=\Sigma(\tau)-c_k\,h\,D^{-1}(\tau)+o_p(h)
\end{displaymath}
by Theorem 3. This means that $\widehat{\Sigma}_h(\tau)$ captures the improvement in precision due to smoothing and, accordingly, the length of a confidence interval based on $\widehat{\Sigma}_h(\tau)$ for any element of $\beta(\tau)$ is asymptotically smaller than the length of a confidence interval based on the naive variance estimator $\tau(1-\tau)\widehat{D}_h^{-1}(\tau)\,\frac{1}{n}\sum_{i=1}^nX_iX_i'\widehat{D}_h^{-1}(\tau)$.

Note that both (S) and (U) make use of the implicit function theorem applied to the first-order condition $\widehat{R}_h^{(1)}\big(\widehat{\beta}_h(\tau);\tau\big)=0$ to obtain
\begin{displaymath}
\frac{\partial\widehat{\beta}_h(\tau)}{\partial h}=-\left[\widehat{R}_h^{(2)}\big(\widehat{\beta}_h(\tau);\tau\big)\right]^{-1}\frac{\partial\widehat{R}_h^{(1)}\big(\widehat{\beta}_h(\tau);\tau\big)}{\partial h}.
\end{displaymath}
We then establish a convergence rate that holds uniformly in $h$ for $\widehat{R}_h^{(2)}\big(\widehat{\beta}_h(\tau);\tau\big)$ and $\frac{\partial}{\partial h}\,\widehat{R}_h^{(1)}\big(\widehat{\beta}_h(\tau);\tau\big)$, leading to
\begin{eqnarray*}
\sqrt{n}\left(\widehat{\beta}_{h_1}(\tau)-\beta_{h_1}(\tau)-\left[\widehat{\beta}_{h_0}(\tau)-\beta_{h_0}(\tau)\right]\right)&=&\sqrt{n}\int_{h_0}^{h_1} \left[\frac{\partial\widehat{\beta}_h(\tau)}{\partial h}-\frac{\partial\beta_h(\tau)}{\partial h}\right]\mathrm{d}h\\
&=&\sqrt{n}\,O_p\left(\int_{h_0}^{h_1}\sqrt{\frac{\ln n}{n h}}\,\mathrm{d}h\right)=O_p(\sqrt{\ln n})\left|\sqrt{h_1}-\sqrt{h_0}\right|,
\end{eqnarray*}
uniformly for $(h_0,h_1)\in[\ubar{h}_n,\overline{h}_n]^2$.

\subsection{Extensions}\label{S:extensions}

\paragraph{Pdf curve estimation.}~~~~The next result states the uniform consistency of the pdf curve estimator. A distinctive feature of the estimation procedure we propose is that consistency holds even when the covariate value $x$ is not in the support of $X$. The key for this ability to extrapolate lies on the underlying quantile regression specification, which we assume correct.\vskip 1em

\noindent\textbf{Proposition 1}~~~~\emph{Under Assumptions~X, Q and K,} $\norm{\widehat{\texttt{f}}_h(\tau|x)-\texttt{f}(\tau|x)}=o(h^s)+O_p\big(\sqrt{\ln n/(nh)}\big)$ \emph{uniformly for $x$ in any compact set, $\tau$ in $[\underline{\tau},\overline{\tau}]$, and $h$ in $[\underline{h}_n,\overline{h}_n]$.}\vskip 1em

Proposition 1 shows that the dimension of the covariate vector $X$ does not affect the order $\sqrt{\ln n/(nh)}$ of the stochastic estimation error of $\widehat{\texttt{f}}_h(\tau|x)$, so that the curse of dimensionality does not apply. This result obviously depends on the correct specification of the quantile regression model. The bias in Proposition~1 is of order $o(h^s)$ due to the Hessian bias in Lemma~1. The latter arises instead of the usual $O(h^s)$ bias order because we calibrate the kernel order for the QR estimation, which involves a function with $s+1$ derivatives. In contrast, it is the estimation of the pdf, which has only $s$ derivatives, that drives the bias in the Hessian term.

As it turns out, the quantile density estimator $\widehat{q}_h (\tau|x)$ drives the nonparametric consistency rate. Alternatively, we could estimate the pdf by $\widehat{f}_h(y|x)=1/\widehat{q}_h\big(\widehat{F}_h(y|x)|x\big)$, where $\widehat{F}_h(y|x)$ is a conditional cdf estimator. The consistency rate of such $\widehat{f}_h(y|x)$ is the same as in Proposition 1 if $\widehat{F}_h(y|x)$ ensues from inverting $x'\widehat{\beta}_h(\tau)$ for $y$ in its range.

\paragraph{Semiparametric efficiency.}~~~~We next establish the asymptotic equivalence of the feasible QR estimator $\check{b}_q(\tau)$ in (\ref{Weighted}) to its unfeasible counterpart $\tilde{b}_q(\tau)$ in (\ref{Infefficient}). In addition, we also document the asymptotic normality and efficiency of both estimators. Before stating the result, recall that \cn{Newey1990efficient} show that $\Sigma_q(\tau):=\tau(1-\tau)D_q^{-1}(\tau)$, with
\begin{displaymath}
D_q(\tau):=\E\left[\frac{XX'}{q^2(\tau|X)}\right]=\E\left\{XX'f^2[X'\beta(\tau)|X]\right\},
\end{displaymath}
is the asymptotic covariance matrix of the asymptotically efficient QR estimator.\vskip 1em

\noindent\textbf{Proposition 2}~~~~\emph{Let $m=o(n^{1/2})$ and $h=o(1)$, with $1/h=o\big((m/\ln m)^{1/3}\big)$. Assumptions~X, Q and K1 ensure that $\sqrt{n}\left(\check{b}_q(\tau)-\tilde{b}_q(\tau)\right)=O_p(\varrho_q+\varrho_s+\varrho_{bk})$, where $\varrho_q=h^s+\sqrt{\ln m/(mh)}$, $\varrho_s=\sqrt{m/n}$, and $\varrho_{bk}=(\ln n)^{3/4}\,n^{-1/4}$. In addition, both $\sqrt{n}\big(\check{b}_q(\tau)-\beta(\tau)\big)$ and $\sqrt{n}\big(\tilde{b}_q(\tau)-\beta(\tau)\big)$ converge in distribution to $\mathcal{N}\big(0,\Sigma_q(\tau)\big)$.}\vskip 1em

Both $\check{b}_q(\tau)$ and $\tilde{b}_q(\tau)$ are asymptotically efficient, with asymptotic covariance matrix $\Sigma_q(\tau)$. To estimate the latter, one may employ a weighted version of the Hessian $\widehat{R}^{(2)}_h (b;\tau)$. The rate at which the difference $\sqrt{n}\left(\check{b}_q(\tau)-\tilde{b}_q(\tau)\right)$ shrinks to zero involves three components. The first is the consistency rate $\varrho_q$ of the qdf estimator $\check{q}_h(\tau|x)$ in Proposition 1. Using standard kernel estimators as in \cn{otsu2008conditional} and \cn{komunjer2010efficient} or $k-$NN smoothing as in \cn{Newey1990efficient} and \cn{Zhao2001asymptotically} would also involve a variance term of order $1/(mh^d)$ due to the curse of dimensionality. The rate $\varrho_s$ comes from the sample splitting scheme, whereas $\varrho_{bk}$ arises from the Bahadur-Kiefer linear approximation.

A natural extension of Proposition 2 is to consider smoothed versions of the objective functions corresponding to the efficient feasible and infeasible estimators, in order to obtain better AMSE performances.

\section{Monte Carlo study}

To assess how well the asymptotic theory reflects the performance of our estimator in finite samples, we run simulations for a median linear regression: $Y=X'\beta+\epsilon$, where $X=(1,\tilde{X})$, with $\tilde{X}\sim U[1,5]$, and $\beta\equiv\beta(1/2)=(1,1)$. We entertain five different specifications for the error distribution. The first three are asymmetric, with errors coming from exponential, Gumbel (type I extreme value), and $\chi_3^2$ distributions. The fourth specification displays heavy tails, with errors following a t-student with 3 degrees of freedom. The fifth distribution exhibits conditional heteroskedasticity: $\epsilon=\frac{1}{4}\,(1+\tilde{X})\,Z$, with $Z\sim\mathcal{N}(0,1)$. We recenter, if necessary, the error distributions to ensure median zero and, except to the $\chi_3^2$ distribution, also rescale them to obtain unconditional variance equal to two. Apart from the exponential distribution, the other specifications are as in \cn{horowitz1998bootstrap}, \cn{whang2006smoothed}, and \cn{kaplan2017smoothed}.

In each of the 100,000 replications, we sample $n\in\{100,250,500,1000\}$ observations and then compute \pc{koenker1978regression} standard median regression estimator (MR), \pc{horowitz1998bootstrap} smoothed median regression estimator (SMR), and our convolution-type kernel estimator (CKMR). We also compute the empirical coverage of their asymptotically-valid confidence intervals at the 95\% and 99\% levels. We gauge the latter as the proportion of replications in which the absolute value of the t-statistic is below the corresponding percentile in the standard normal distribution (namely, 1.96 and 2.58 at the 95\% and 99\% confidence levels, respectively).

We compute the smoothed estimators using a standard Gaussian kernel and a bandwidth grid with values ranging from $0.08$ to $0.80$, with increments of 0.02. In addition, we also evaluate them at \pc{silverman1986density} rule-of-thumb bandwidth $h_{ROT}=1.06\,\widehat{s}\,n^{-1/5}$, where $\widehat{s}$ is the minimum between the sample standard deviation and the interquartile range (divided by 1.38898) of the standard MR residuals. We compute the MR standard errors by a pairs bootstrap procedure as usual in the literature,\footnote{~~We omit results using standard errors as in \cn[Sections 3.4.2 and 4.10.1]{koenker2005quantile} because bootstrapping always entails better empirical coverage. They are available from the authors upon request.} whereas the SMR standard errors are as in \cn[Section 2]{horowitz1998bootstrap}. Finally, we estimate the standard errors for the CKMR estimator using the square root of the diagonal entries of $\widehat{\Sigma}_h(1/2)$.

Figures \ref{fig:exponential} to \ref{fig:heteroskedastic} show how each estimator performs across distributions and sample sizes. For simplicity, we focus exclusively on the slope coefficient given that we find little difference between the intercept estimates. For every figure, the first row displays the relative mean squared error (RMSE) of the smoothed estimators with respect to the standard MR estimator across different sample sizes, whereas the second row depicts the standard error of the slope estimators, within their one-standard-deviation band. The third and fourth rows portray respectively the empirical coverage of the asymptotic confidence intervals at the 95\% and 99\% levels.

The smoothed estimators obviously depend on the bandwidth choice, and hence we plot their results as a function of the bandwidth $h$, singling out the corresponding outcomes for the rule-of-thumb bandwidth $h_{ROT}$. As the value of the latter changes across replications, we arbitrarily place the results at the average value of $h_{ROT}$ across the 100,000 replications. The inside tick marks in the horizontal axis depict the deciles of the distribution of the rule-of-thumb bandwidth across replications. The distribution seems symmetric in every instance, with dispersion reducing drastically once we move from a sample size of 100 to 1,000 observations.

Smoothing seems to pay off as both SMR and CKMR entail lower MSE than the standard MR estimator for every distribution. Interestingly, the RMSE of the CKMR estimator is less sensitive to the sample size and bandwidth value, especially around the rule-of-thumb choice, than the RMSE of the SMR estimator. For instance, if errors are exponential as in Figure~\ref{fig:exponential}, the CKMR estimator using the rule-of-thumb bandwidth yields a RMSE of about 70\% for the smaller sample sizes, and between 75\% and 80\% for the sample sizes of 500 and 1,000 observations. Conversely, the RMSE of the SMR estimator changes from 80\% to 90\% as we increase the sample size from 100 to 250 observations and then cease to improve the standard MR estimator in terms of MSE for the larger sample sizes.

Regardless of the specification we consider, the key to the CKMR success lies on the much lower standard errors. It is apparent from Figures~\ref{fig:exponential}~to~\ref{fig:heteroskedastic} that the standard errors of the SMR estimator that uses the rule-of-thumb bandwidth are very close to the bootstrap-based standard errors of the MR estimator. They are slightly lower for the exponential and chi-squared errors, but slightly larger for the Gumbel, t-student and heteroskedastic specifications.

The confidence intervals match very well their nominal values for every estimator at the 95\% and 99\% levels. This is particularly reassuring for the CKMR estimator, whose confidence intervals are significantly narrower due to the smaller standard errors. In addition, the empirical coverage of the CKMR estimator is very stable across bandwidth values, especially for samples with 250 or more observations. In contrast, the empirical coverage of the SMR estimator deteriorates for larger bandwidth values regardless of the error distribution and sample size. All in all, the CKMR estimator performs very well relative to the MR and SMR estimators, with lower MSE and tighter (asymptotic) confidence intervals that are nonetheless valid even in small samples.

\subsection{Conditional qdf and pdf estimation}

Next, we complement the discussion in Section \ref{sec:pdf-estimator} by illustrating the qdf and pdf estimation of the QR model $Y=\beta_0(U)+\beta_1(U)X_1+\beta_2(U)X_2+\beta_3(U)X_3$, where $U$, $X_1$, $X_2$ and $X_3$ are independent and uniformly distributed variables on the unit interval $[0,1]$, $\beta_0(\cdot)$ and $\beta_1(\cdot)$ are respectively given by the $\mathsf{Beta}(1,16)$ and $\mathsf{Beta}(32,32)$ quantile functions, $\beta_2(\tau)=1$, and $\beta_3(\tau)=\frac{(2\pi+8)\tau-(\cos(2\pi\tau)-1)}{2\pi+8}$. The large number of regressors relative to the sample size of 1,000 observations does not bode well for standard nonparametric estimation due to the curse of dimensionality.

In each of the 1,000 replications, we estimate the pdf by means of \eqref{eq:density-estimator} using the rule-of-thumb bandwidth $h_{ROT}$.\footnote{~~It is perhaps worth mentioning that the optimal bandwidth for the QR estimation is not necessarily optimal for the qdf estimator.} Figure \ref{fig:pdf-estimator-simu} displays the true quantile function in Panel (a), the true qdf in Panel (b), and the true conditional density of $Y$ given $(x_1,x_2,x_3)=(0.9,0.5,0.9)$ in Panel (c). The gray-shaded regions in Panels (a) and (b) correspond to the interquantile band from 0.05 to 0.95, i.e., the region for which at most 5\% of the replications lay below its lower bound and at most 5\% above its upper bound, at each fixed $\tau$.

Figure \ref{fig:pdf-estimator-simu}(a) suggests that the variability and bias of the QR estimator are not only quite small but also of a similar magnitude. Panel (b) reveals that the qdf estimates approximate well the poles in the tails, up to some boundary bias at $\tau=0$, with a very narrow band regardless of the quantile level $\tau$. Panel (c) displays the \emph{paths} $\tau\mapsto\widehat{\texttt{f}}_h(\tau\vert x)$ for $\tau\in\{0.01,0.02,\dots,0.99\}$, corresponding to the 100 first realizations of $\widehat{\texttt{f}}_h$. We find a much higher variability for the pdf curve estimates, as typical in nonparametric estimation. As before, we evince some boundary bias at $\tau=0$ and tighter interquantile bands in the tails. Altogether, the pdf estimator seems to perform reasonably well even for three conditioning variables.

\section{Concluding remarks}

This paper proposes a convolution-type kernel QR estimator based upon smoothing the objective function. The resulting estimator improves on standard QR estimation both in terms of asymptotic mean square error (MSE) and Bahadur-Kiefer representation. Applying the implicit function theorem to the corresponding first-order condition gives way to a conditional pdf estimator that avoids the curse of dimensionality by exploiting the QR specification. We also show how to use this pdf estimator to come up with a two-step efficient QR estimator. Many of our asymptotic results are uniform with respect to the quantile level and bandwidth. The latter is convenient because it makes room for data-driven bandwidth choices, such as the rule-of-thumb bandwidth we propose. In particular, we establish a functional central limit theorem for our smoothed QR estimator that accommodates data-driven bandwidths.

Simulations show that the rule-of-thumb bandwidth works well in practice, yielding very good results in terms of confidence interval coverage. They also reveal that our smoothed QR estimator outperforms not only the standard QR estimator, but also \pc{horowitz1998bootstrap} alternative smoothed estimator. Finally, our Monte Carlo study also confirms that our QR-based pdf estimator does not suffer from the curse of dimensionality.

There are many extensions to consider in future research. First, uniformity in bandwidth is useful not only for adaptive bandwidth choices as in \cn{lepski1997optimal}, but also for bandwidth-snooping-robust inference as in \cn{AK18}. Second, one could combine our QR-based pdf estimator in a data-rich environment with \pc{belloni2011penalized} lasso-type methods to select the relevant covariates and with \pc{volgushev2019distributed} to cope with large sample sizes. Third, one could exploit the connection between QR and quantile instrumental variable models \cite{kaido2018decentralization} to obtain smoother estimators and better inferential procedures \cite{chernozhukov2005ivqte,J08,AM16,dCGK2017}. Finally, one could also extend our smoothing approach to handle panel QR as in \cn{GK2016}.

\lineskip=.35ex \baselineskip 3.5ex
\bibliography{Quantiles}

\newpage
\appendix\section{Proofs of main results}

Notice that $\widehat{R}(b;\tau)$ is integrable if and only if $Y$ and $X$ are integrable. This matters for the definition of $\beta(\tau)$ as the minimizer of $R(b;\tau):=\E\big[\widehat{R}(b;\tau)\big]$. It is convenient to assume that both $\widehat{R}(b;\tau) $ and $\widehat{R}_h(b;\tau)$ are integrable such that $R(b;\tau)$ and $R_h(b;\tau)$ are well defined. Otherwise, one should define $R_h(b;\tau)$ as $\E\big[\widehat{R}_h(b;\tau)-\widehat{R}_h(0;\tau)\big]$, and similarly for $R (b;\tau)$. These quantities are finite under Assumption~X. Let $\mathcal{S}$ denote the set $\mathbb{R}^d\times[\ubar{\tau},\bar{\tau}]\times [\ubar{h}_n,\bar{h}_n]$ to which $(b,\tau,h)$ belongs. Note that $\mathcal{S}$ depends on $n$ through $[\ubar{h}_n,\bar{h}_n]$. In what follows, $C$ denotes a generic constant that may vary from line to line. Lastly, the online appendix collects the proofs of the intermediary results.\vskip 1em

\subsection{Smoothing bias}

To study the bias of our smoothed QR estimator, we make use of the following result.\vskip 1em

\noindent\textbf{Lemma 1}~~~~\emph{Assumptions~X, Q2 and K1 ensure that}
\begin{eqnarray*}
(i)\phantom{iiv}&&\displaystyle{\sup_{(b,\tau,h)\in\mathcal{S}}\abs{\frac{R_h(b;\tau)-R(b;\tau)}{h^{s+1}}}=O(1)};\\
(ii)\phantom{iv}&&\displaystyle{\sup_{(b,\tau,h)\in\mathcal{S}}\norm{\frac{R_h^{(1)}(b;\tau)-R^{(1)}(b;\tau)}{h^{s+1}}}=O(1)};\\
(iii)\phantom{v}&&\displaystyle{\sup_{(b,\tau,h)\in\mathcal{S}}\norm{\frac{R_h^{(2)}(b;\tau)-R^{(2)}(b;\tau)}{h^s}}=o(1)};\\
(iv)\phantom{ii}&&\displaystyle{\sup_{(\delta,b,\tau,h)\in\mathbb{R}^d\times\mathcal{S}}\norm{\frac{R_h^{(2)}(b+\delta;\tau)-R_h^{(2)}(b;\tau)}{\norm{\delta}}}=O(1)}.
\end{eqnarray*}
\noindent\textbf{Proof of Lemma~1}~~~~See online appendix$.\hfill\blacksquare$\vskip 1em

\noindent\textbf{Proof of Theorem~1}~~~~Observe first that $R(\cdot;\tau)$ is strictly convex. To see this, we show that, for any $b\in\mathbb{R}^d$, $R^{(2)}(b;\tau)$ is an element of the open set $M_d^+$ of positive-definite $d\times d$ matrices. Indeed, for any $v\in\mathbb{R}^d$,
\begin{displaymath}
v'R^{(2)}(b;\tau)v=\int(v'x)^2f(x'b\given x)\,\mathrm{d}F_X(x)\ge v'\E(XX')v\,\inf_{x\in\supp(X)}f(x'b\given x)>0
\end{displaymath}
by Assumptions~X and Q. It follows from Lemma 1($i$) and the continuity of $b\mapsto R^{(2)}(b;\tau)=R^{(2)}(b)$ that any potential minimizer $\beta_h(\tau)$ of $R_h(\cdot;\tau)$ must lie in a compact $\mathcal{B}$ for every admissible $(\tau,h)$. As Lemma 1($i$) ensures that $R_h (\cdot;\tau)$ is convex over $\mathcal{B}$ for all admissible $\tau$ and $h$ small enough, it follows that $\beta_h(\tau)$ is unique for $(h,\tau)\in[\ubar{h}_n,\bar{h}_n]\times[\ubar{\tau},\bar{\tau}]$ and $n$ large enough.

By the implicit function theorem, $\beta_h(\tau)$ is continuous in $(\tau,h)$ given that $R_h^{(1)}(b;\tau)$ and $R_h^{(2)}(b;\tau)$ are continuous in $(b,\tau,h)$ and that $R_h^{(1)}(\beta_h(\tau);\tau)=0$. Let
\begin{displaymath}
A(\tau,h):=\int_0^1 R^{(2)}\Big(\beta(\tau)+w\big[\beta_h(\tau)-\beta(\tau)\big];\,\tau\Big)\,\mathrm{d}w
\end{displaymath}
and let $c:=\inf f\Big(x'\big[\beta(\tau)+w\big(\beta_h(\tau)-\beta(\tau)\big)\big]\given x\Big)$ for $(\tau,h,w,x)\in[\ubar{\tau},\bar{\tau}]\times[\ubar{h}_n,\bar{h}_n]\times[0,1]\times\supp(X)$. Assumptions~X and Q ensure that $c>0$, and so
\begin{displaymath}
v'A(\tau,h)v=\int_0^1\int(v'x)^2f\Big(x'\big[\beta(\tau)+w\big(\beta_h(\tau)-\beta(\tau)\big)\big]\given x\Big)\,\mathrm{d}F_X(x)\,\mathrm{d}w\ge c\,v'\E(XX')v>0
\end{displaymath}
by Assumption~X. This means that the eigenvalues of $A(\tau,h)$ are bounded away from zero, uniformly for $(\tau,h)\in[\ubar{\tau},\bar{\tau}]\times[\ubar{h}_n,\bar{h}_n]$. However, as $R^{(1)}_h\big(\beta_h(\tau);\tau\big)=R^{(1)}\big(\beta(\tau);\tau\big)=0$, a Taylor expansion with integral remainder leads to
\begin{displaymath}
R^{(1)}\big(\beta_h(\tau);\tau\big)-R_h^{(1)}\big(\beta_h(\tau);\tau\big)=R^{(1)}\big(\beta_h(\tau);\tau\big)-R^{(1)}\big(\beta(\tau);\tau\big)=A(\tau,h)\big[\beta_h(\tau)-\beta(\tau)\big],
\end{displaymath}
and hence, by Lemma~1($ii$), $\beta_h(\tau)-\beta(\tau)=O(h^{s+1})$ given that
\begin{displaymath}
\norm{\frac{\beta_h(\tau)-\beta(\tau)}{h^{s+1}}}\le\sup_{(\tau,h)\in[\ubar{\tau},\bar{\tau}]\times[\ubar{h}_n,\bar{h}_n]}\left\{\norm{A^{-1}(\tau,h)}\norm{\frac{R^{(1)}\big(\beta_h(\tau);\tau\big) -R^{(1)}_h\big(\beta_h(\tau);\tau\big)}{h^{s+1}}}\right\}=O(1).
\end{displaymath}
Now, a Taylor expansion obtains
\begin{displaymath}
-R_h^{(1)}\big(\beta(\tau);\tau\big)=R_h^{(1)}\big(\beta_h(\tau);\tau\big)-R_h^{(1)}\big(\beta(\tau);\tau\big)=\left[R^{(2)}\big(\beta(\tau);\tau\big)+o(1)\right]\!\left[\beta_h(\tau)-\beta(\tau)\right]+o(h^{s+1})
\end{displaymath}
by Lemma~1. It follows from $\int_0^1 w(1-w)^{s-1}\,\mathrm{d}w=\frac{1}{(s+1)s}$ and \eqref{eq:i1} that
\begin{displaymath}
T_2:=\int_{-\infty}^{x'\beta(\tau)}\left\{\E\big[k_h(v-Y)\given x\big]-f(v\given x)\right\}\mathrm{d}v=h^{s+1}\frac{\int z^{s+1}\,k(z)\,\mathrm{d}z}{(s+1)!}\,f^{(s)}(x'\beta(\tau)\given x)+o(h^{s+1}).
\end{displaymath}
Using \eqref{eq:R_h-gradient-representation} then yields
\begin{displaymath}
R_h^{(1)}\big(\beta(\tau);\tau\big)=h^{s+1}\frac{\int z^{s+1}\,k(z)\,\mathrm{d}z}{(s+1)!}\,\int x\,f^{(s)}\left(x'\beta(\tau)\given x\right)\,\mathrm{d}F_X(x) +o(h^{s+1}),
\end{displaymath}
completing the proof$.\hfill\blacksquare$\vskip 1em

\subsection{Bahadur-Kiefer representation}

This section makes use of a powerful functional exponential inequality by \cn{massart2007concentration}. For the sake of completeness, we state a version of \cnm{massart2007concentration}'s result as a Lemma. For real-valued functions $\ubar{f}$ and $\bar{f}$ with $\ubar{f}\le\bar{f}$, let $[\ubar{f},\bar{f}]$ denote the set of all functions $g$ such that $\ubar{f}\le g\le\bar{f}$. For a set $\mathcal{F}$ and a family $\{F_i\}$ of subsets of $\mathcal{F}$, we say that $\{F_i\}$ \emph{covers} $\mathcal{F}$ if $\mathcal{F}\subset\bigcup_i F_i$.\vskip 1em

\noindent\textbf{Lemma~2}~~~~\emph{Let $Z_i$ be an iid sequence of random variables taking values in the measurable space $\mathcal{Z}$, and let $\mathcal{F}$ be a class of real valued, measurable functions on $\mathcal{Z}$. Assume that
\begin{description}
\item[$(i)\phantom{i}~$] there are some positive constants $\sigma$ and $M$ such that $\E\!\left[\abs{f(Z_i)}^2\right]\le\sigma^2$ and $\sup_{z\in\mathcal{Z}}\abs{f(z)}\le M$ for all $f\in\mathcal{F}$;
\item[$(ii)~$]for each $\delta>0$, there exists a set of brackets $\big\{[\ubar{f}_j,\bar{f}_j];\;j=1,\dots,J(\delta)\big\}$, for some integer $J(\delta)>1$, that covers $\mathcal{F}$ such that $\E\!\left[\abs{\bar{f}_j(Z_i)-\ubar{f}_j(Z_i)}^2\right]\le\delta^2$ and $\sup_{z\in\mathcal{Z}} \abs{\bar{f}_j(z)-\ubar{f}_j(z)}\le M$ for every $j=1,\ldots,J(\delta)$.
\end{description}
It then follows that, for any $r\ge 0$,
\begin{equation}\label{eq:maximal-inequality}
\Prob\left(\sup_{f\in\mathcal{F}}\frac{1}{\sqrt{n}}\sum_{i=1}^n\left(f\!(Z_i)-\E[f(Z_i)]\right)\ge\mathcal{H}_n+7\sigma\sqrt{2r}+\frac{2Mr}{\sqrt{n}}\right)\le\exp(-r),
\end{equation}
where
\begin{displaymath}
\mathcal{H}_n := 27\left(\int_0^\sigma H^{1/2}(u)\,\mathrm{d}u+\frac{2(\sigma+M)H(\sigma)}{\sqrt{n}}\right)
\end{displaymath}
and $H$ is any nonnegative measurable function of $\delta>0$ satisfying $H(\delta)\ge\ln J(\delta)$.}\\
\noindent\textbf{Proof of Lemma 2}~~~~See Corollary 6.9 in \cn{massart2007concentration}$.\hfill\blacksquare$\vskip 1em

Before proceeding to the next result, let us introduce some additional notation. In what follows, let $\sup_{(\tau,h)}$ denote the supremum over $(\tau,h)\in[\ubar{\tau},\bar{\tau}]\times[\ubar{h}_n,\bar{h}_n]$, which depends on $n$ via $[\ubar{h}_n,\bar{h}_n]$. We also use the same implicit notation for any other similar operator (e.g., infimum or union). Recalling that $\varrho_n^{-1}(h):=\sqrt{nh/(\ln n)}$, let
\begin{displaymath}
\calE:=\left\{\sqrt{n}\sup_{(\tau,h)}\varrho_n^{-1}(h)\norm{\widehat{\beta}_h(\tau)-\beta_h(\tau)+D_h^{-1}(\tau)\widehat{S}_h(\tau)}\ge r^2\right\}.
\end{displaymath}
Notice that the event $\calE$ depends on the sample size $n$ and on a tail parameter $r$, but neither on $\tau$ nor on $h$. On the complementary set of $\calE$, it holds that
\begin{displaymath}
\sqrt{n}\big(\widehat{\beta}_h(\tau)-\beta_h(\tau)\big)=-\sqrt{n}\,D_h^{-1}(\tau)\widehat{S}_h(\tau)+\widehat{E}_h(\tau),
\end{displaymath}
where the approximation error is such that $\big\Vert{\widehat{E}_h(\tau)\big\Vert}\le\varrho_n(h)\,r^2$ uniformly for $(\tau,h)\in[\ubar{\tau},\bar{\tau}]\times[\ubar{h}_n,\bar{h}_n]$. In particular, if $\Prob\big(\calE\big)$ is small for large $r$, then the representation~\eqref{eq:bahadur-representation} from Theorem~2 holds uniformly in $\tau$ and $h$.

Finally, let $\breve{A}$ denote the complementary set of $A$ and
\begin{eqnarray*}
\calEone&:=&\left\{\sup_{(\tau,h)}\norm{\sqrt{n}\,\widehat{S}_h(\tau)}\ge r\right\}, \\
\calEtwo&:=&\left\{\sup_{(\tau,h)}\sup_{\{b:\,\norm{b-\beta_h(\tau)}\le 1\}}\norm{\sqrt{\frac{nh}{\ln n}}\left(\widehat{R}_h^{(2)}(b;\tau)- R_h^{(2)}(b;\tau)\right)}\ge r\right\},
\end{eqnarray*}
where the norms in $\calEone$ and $\calEtwo$ are the Euclidean and operator norms, respectively. We are now ready to state the functional exponential inequality we will apply in the remaining technical proofs.\vskip 1em

\noindent\textbf{Proposition 3}~~~~\emph{Given Assumptions~X, Q and K, $\widehat{\beta}_h(\cdot)$ is unique and continuous over $[\ubar{\tau},\bar{\tau}]\times[\ubar{h}_n,\bar{h}_n]$ with probability growing to one. There also exist positive constants $C_0$, $C_1$ and $C_2$ such that, for small enough $\epsilon$, $1/r$ and $1/n$,}
\begin{eqnarray*}
(i)\phantom{iiv}&&\displaystyle{\Prob\left(\calE\cap\calEonec\cap\calEtwoc\right)\le C_0\exp\left(-n\epsilon/C_0\right)};\\
(ii)\phantom{iv}&&\displaystyle{\Prob\left(\calEone\right)\le C_1\exp\left(-r^2/C_1\right)};\\
(iii)\phantom{v}&&\displaystyle{\Prob\left(\calEtwo\right)\le C_2\exp\left(-r\ln n/C_2\right)};
\end{eqnarray*}
\textbf{Proof of Proposition 3}~~~~See online appendix$.\hfill\blacksquare$\vskip 1em

Theorem~2 is an immediate corollary to Proposition~3, whose proof builds on Lemmata~2~to~4. It is sometimes convenient to consider the auxiliary objective functions $\widehat{\mathcal{R}}_h(b;\tau):=\widehat{R}_h(b;\tau)-\widehat{R}_h\big(\beta_h(\tau);\tau\big)$ and $\mathcal{R}_h(b;\tau):=\E\big[\widehat{\mathcal{R}}_h(b;\tau)\big]$. They are such that $\widehat{\beta}_h(\tau)=\arg\min_b\widehat{\mathcal{R}}_h(b;\tau)$ and $\beta_h(\tau)=\arg\min_b\mathcal{R}_h(b;\tau)$. Similarly, set $\widehat{\mathcal{R}}(b;\tau)=\widehat{R}(b;\tau)-\widehat{R}\big(\beta(\tau);\tau\big)$ and $\mathcal{R}(b;\tau):=\E\big[\widehat{\mathcal{R}}(b;\tau)\big]$. The next result shows that $\widehat{\beta}_h(\tau)$ is close to $\beta_h(\tau)$ uniformly for $(\tau,h)\in[\ubar{\tau},\bar{\tau}]\times[\ubar{h}_n,\bar{h}_n]$.\vskip 1em

\noindent\textbf{Lemma 3}~~~~\emph{Suppose that Assumptions~X, Q and K hold. If $n$ is large enough, there is a positive constant $C$ such that, for any $\eta\in[1/\ln n,1]$,}
\begin{displaymath}
\Prob\left(\sup_{(\tau,h)}\norm{\widehat{\beta}_h(\tau)-\beta_h(\tau)}\ge\eta\right)\le C\exp(-C\,n\,\eta^4).
\end{displaymath}
\textbf{Proof of Lemma 3}~~~~See online appendix$.\hfill\blacksquare$\vskip 1em

\noindent\textbf{Lemma 4}~~~~\emph{Suppose that Assumptions~X, Q, and K hold and consider $r>0$ and $\eta\in(0,1]$. As long as $n$ is large enough,
\begin{eqnarray*}
\Prob\left(\sup_{(\tau,h)}\norm{\sqrt{n}\widehat{S}_h(\tau)}\ge C_1(1+r)\right)&\le&C_0\,\exp\left(-r^2\right),\\
\Prob\left(\sup_{(\tau,h)}\sup_{\big\{b:\,\norm{b-\beta_h(\tau)}\le\eta\big\}}\norm{\sqrt{\frac{nh}{\ln n}}\left(\widehat{R}_h^{(2)}(b,\tau)- R_h^{(2)}(b,\tau)\right)}\ge C_1(1+r)\right)&\le&C_0\,\exp\left(-r\ln n\right).
\end{eqnarray*}
Moreover, for any compact subset $\mathcal{B}$ of $\mathbb{R}^{d+1}$,}
\begin{eqnarray*}
\sup_{(b_0,b_1)\in\mathcal{B}^2}\sup_{(\tau,h)}\frac{\norm{\widehat{R}_h^{(2)}(b_1,\tau)-\widehat{R}_h^{(2)}(b_0,\tau)}}{\norm{b_1-b_0}}&=&O_p (1),\\
\sup_{(b_0,b_1)\in\mathcal{B}^2}\sup_{(\tau,h)}\frac{\norm{\frac{1}{n}\sum_{i=1}^nX_iX_i'\left\{\left[K\big(\!-\!e_i(b_1)/h\big)-\tau\right]^2-\left[K\big(\!-\!e_i(b_0)/h\big)-\tau\right]^2\right\}}}{\norm{b_1-b_0}}&=&O_p(1),\\
\sup_{b\in\mathcal{B}}\sup_{(\tau,h)}\norm{\frac{1}{n}\sum_{i=1}^nX_iX_i'\left[K\big(\!-\!e_i(b)/h\big)-\tau\right]^2-\E\!\left[X_iX_i'\left[K\big(\!-\!e_i(b)/h\big)-\tau\right]^2\right]}&=&O_p(1/\sqrt{n}).
\end{eqnarray*}
\textbf{Proof of Lemma 4}~~~~See online appendix$.\hfill\blacksquare$\vskip 1em

\noindent\textbf{Proof of Theorem~2}~~~~As uniqueness holds, to establish \eqref{eq:bahadur-representation}, it suffices to fix $\epsilon$ and increase $r$ to make $\Prob\left(\calE\right)\le\Prob\left(\calE\cap\Big[\calEonec\cap\calEtwoc\Big]\right)+\Prob\left(\calEone\right)+\Prob\left(\calEtwo\right)$ arbitrarily small for $n$ large enough. As for \eqref{eq:nonlinear-representation}, it readily follows from \eqref{eq:integral-representation} in the online appendix and Lemma~4 in view that the latter and \eqref{eq:bahadur-representation} imply that $\sup_{\tau,h}\norm{\widehat{\beta}_h(\tau)-\beta_h(\tau)}=O_p(1/\sqrt{n}).\hfill\blacksquare$\vskip 1em

\subsection{Asymptotic variance and mean squared error}

\noindent\textbf{Proof of Theorem~3}~~~~We first show that the expansion
\begin{equation}\label{eq:asymptotic-variance-old}
\V\!\left(\sqrt{n}\,D^{-1}(\tau)\widehat{S}_h(\tau)\right)=\Sigma(\tau)-c_k\,h\,D^{-1}(\tau)+O(h^2)
\end{equation}
holds uniformly with respect to $(\tau,h)\in[\ubar{\tau},\bar{\tau}]\times[\ubar{h}_n,\bar{h}_n]$. Given that $\E\!\left[\widehat{R}_h^{(1)}\big(\beta_h(\tau);\tau\big)\right]=0$,
\begin{eqnarray*}
\V\big(\sqrt{n}\,\widehat{S}_h(\tau)\big)&=&\V\!\left(\sqrt{n}\,\widehat{R}_h^{(1)}\big(\beta_h(\tau);\tau\big)\right)\\
                                         &=&\V\!\left(X\left[K\left(\frac{X'\beta_h(\tau)-Y}{h}\right)-\tau\right]\right)=\E\!\left(XX'\left[K\left(\frac{X'\beta_h(\tau)-Y}{h}\right)-\tau\right]^2\right)\\
                                         &=&\E\!\left[XX'K^2\left(\frac{X'\beta_h(\tau)-Y}{h}\right)\right]-2\,\tau\,\E\!\left[XX'K\left(\frac{X'\beta_h(\tau)-Y}{h}\right)\right]+\tau^2\,\E(XX').
\end{eqnarray*}
Along similar lines to the proof of Lemma~1, it follows from Assumptions~Q2 and K that
\begin{eqnarray}
\E\!\left[\left.K\left(\frac{X'\beta_h(\tau)-Y}{h}\right)\right|X=x\right]&=&\int K\left(\frac{x'\beta_h(\tau)-y}{h}\right)f(y\given x)\,\mathrm{d}y=\frac{1}{h}\int k\left(\frac{x'\beta_h(\tau)-y}{h}\right)F(y\given x)\,\mathrm{d}y\nonumber\\
                                                                  &=&F\big(x'\beta_h(\tau)\given x\big)+\int\left[F\big(x'\beta_h(\tau)-hz\given x\big)-F\big(x'\beta_h(\tau)\given x\big)\right]k(z)\,\mathrm{d}z\nonumber\\
                                                                  &=&\tau+O(h^{s+1}), \label{eq:thm-avar-order-1}
\end{eqnarray}
using integration by parts and Theorem~1, in view that $x'\beta(\tau)=F^{-1}(\tau\given x)$ by definition. Let now $\mathbf{K}(z)=2\,k(z)\,K(z)=\frac{\mathrm{d}}{\mathrm{d}z}\,K^2(z)$, so that $\int\mathbf{K}(z)\,\mathrm{d}z= \lim_{z\rightarrow\infty}K^2(z)=1$. As before, this leads to
\begin{eqnarray}
\E\!\left[\left.K^2\left(\frac{X'\beta_h(\tau)-Y}{h}\right)\right|X=x\right]&=&\frac{1}{h}\int\mathbf{K}\left(\frac{x'\beta_h(\tau)-y}{h}\right)F(y\given x)\,\mathrm{d}y\nonumber\\
                                                                            &=&\tau+O(h^{s+1})+\int\left[F\left(x'\beta_h(\tau)-hz\given x\right)-F\left(x'\beta_h(\tau)\given x\right)\right]\mathbf{K}(z)\,\mathrm{d}z\nonumber\\
                                                                            &=&\tau+O(h^{s+1})-h\left[f\big(x'\beta_h(\tau)\given x\big)+O(h)\right]\int z\,\mathbf{K}(z)\,\mathrm{d}z\nonumber\\
                                                                            &=&\tau+O(h^{s+1})-h\left[f\big(x'\beta(\tau)\given x\big)+O(h^{(s+1)})+O(h)\right]\int z\,\mathbf{K}(z)\,\mathrm{d}z\nonumber\\
                                                                            &=&\tau-h\,f\big(x'\beta(\tau)\given x\big)\int z\,\mathbf{K}(z)\,\mathrm{d}z+O(h^2). \label{eq:thm-avar-order-2}
\end{eqnarray}
The variance expansion \eqref{eq:asymptotic-variance-old} then follows by noticing that, as $K\left(-z\right)=1-K(z)$,
\begin{eqnarray*}
\int z\,\mathbf{K}(z)\,\mathrm{d}z&=&2\int z\,k(z)\,K(z)\,\mathrm{d}z=\int_{-\infty}^0 z\,\mathrm{d}K^2(z)+\int_0^\infty z\,\mathrm{d}\left[K^2(z)-1\right]\\
                              &=&-\int_{-\infty}^0K^2(z)\,\mathrm{d}z+\int_0^\infty\left[1-K^2(z)\right]\mathrm{d}z=\int_0^\infty\left\{-\big[1-K(z)\big]^2+1-K^2(z)\right\}\mathrm{d}z\\
                              &=&2\int_0^\infty K(z)[1-K(z)]\,\mathrm{d}z.
\end{eqnarray*}
Finally, the local Lipschitz property of matrix inversion ensures that
\begin{eqnarray*}
\V\!\left(\sqrt{n}\,D_h^{-1}(\tau)\widehat{S}_h(\tau)\right)&=&\Sigma(\tau)-c_k\,h\,D^{-1}(\tau)+O(h^2)\\
                                                            &&+\;\left[D_h^{-1}(\tau)\V\big(\sqrt{n}\,\widehat{S}_h(\tau)\big)D_h^{-1}(\tau)-D^{-1}(\tau)\V\big(\sqrt{n}\,\widehat{S}_h(\tau)\big)D^{-1}(\tau)\right].
\end{eqnarray*}
given \eqref{eq:asymptotic-variance-old}, Lemma~1, and Theorem~1. To establish \eqref{eq:asymptotic-variance}, it suffices to observe that, by Lemma 1, the norm of the last term within brackets is at most equal to
\begin{displaymath}
\norm{D_h^{-1}(\tau)-D^{-1}(\tau)}\norm{\V\big(\sqrt{n}\,\widehat{S}_h(\tau)\big)}\left(\norm{D_h^{-1}(\tau)}+\norm{D^{-1}(\tau)}\right)\le C\norm{D_h^{-1}(\tau)-D^{-1}(\tau)}=o(h).\hfill\blacksquare
\end{displaymath}\vskip 1em

\noindent\textbf{Proof of Theorem~4}~~~~It follows from Theorem~1, \eqref{eq:asymptotic-variance} and $\E\big[\widehat{S}_h(\tau)\big]=0$ that
\begin{displaymath}
\AMSE\big(\lambda'\widehat{\beta}_h(\tau)\big)=h^{2s+2}\left[\lambda'B(\tau)\right]^2+\frac{1}{n}\,\lambda'\left(\Sigma(\tau)-c_k\,h\,D^{-1}(\tau)\right)\lambda +\underbrace{O(h^2)+o(h^s)}_{=o(h)}.
\end{displaymath}
Letting $g(h)=h^{2s+2}\left[\lambda'B(\tau)\right]^2-n^{-1}c_k\,h\,\lambda'D^{-1}(\tau)\lambda$ and then differentiating with respect to $h$ yields
\begin{displaymath}
g'(h)=(2s+2)\,h^{2s+1}\left[\lambda'B(\tau)\right]^2-\frac{1}{n}\,c_k\,\lambda'D^{-1}(\tau)\lambda.
\end{displaymath}
Solving for $h_\lambda^*$ such that $g'(h_\lambda^*)=0$ yields the desired result and AMSE expansion$.\hfill\blacksquare$\vskip 1em

\subsection{Weak convergence and asymptotic covariance estimator}

In this section, we first establish the weak convergence of $\big\{\sqrt{n}\widehat{S}_h:\,\tau\in[\ubar{\tau},\bar{\tau}]\big\}$ to a Gaussian process in Lemma~5 and then derive the results (G), (V) and (U) in Theorem~5.\vskip 1em

\noindent\textbf{Lemma~5}~~~~\emph{Under Assumptions~X, Q and K, $\big\{\sqrt{n}\widehat{S}_h:\,\tau\in[\ubar{\tau},\bar{\tau}]\big\}$ converges in distribution to a tight, centered Gaussian process $\big\{W_*(\tau):\,\tau\in[\ubar{\tau},\bar{\tau}]\big\}$ with covariance structure}
\begin{equation}\label{eq:asymptotic-gaussian-covariance}
	\V\big(W_*(\tau),W_*(\varsigma)\big)=(\tau\wedge\varsigma-\tau\varsigma)\,\E(XX'),\qquad\tau,\varsigma\in[\ubar{\tau},\bar{\tau}],
\end{equation}
\textbf{Proof of Lemma 5}~~~~See online appendix$.\hfill\blacksquare$\vskip 1em

\noindent\textbf{Proof of Theorem~5}~~~~(G) is a corollary to Theorem 2 and Lemma 5, whereas (V) directly follows from the fact that $\widehat{\beta}_h(\tau)=\beta_h(\tau)+O_p(n^{-1/2})$ uniformly and from Lemma 4. To establish (U), recall that $\widehat{\mathcal{R}}_h^{(1)}\big(\widehat{\beta}_h(\tau);\tau\big)=\mathcal{R}_h^{(1)}\big(\beta_h(\tau);\tau\big)=0$ and $\widehat{\beta}_h(\tau)=\beta_h(\tau)+O_p(n^{-1/2})$ uniformly. By the implicit function theorem, Lemmata 1 and 4 yield that, with probability tending to one,
\begin{eqnarray*}
\frac{\partial\widehat{\beta}_h}{\partial h}&=&-\left[\widehat{\mathcal{R}}_h^{(2)}\big(\widehat{\beta}_h(\tau);\tau\big)\right]^{-1}
\frac{\partial}{\partial h}\,\widehat{\mathcal{R}}_h^{(1)}\big(\widehat{\beta}_h(\tau);\tau\big)=-\left[\widehat{\mathcal{R}}_h^{(2)}\big(\widehat{\beta}_h(\tau);\tau\big)\right]^{-1} \frac{1}{nh}\sum_{i=1}^nX_i\,\tilde{k}\!\left(-\frac{e_i\big(\widehat{\beta}_h(\tau)\big)}{h}\right),\\
\frac{\partial\beta_h}{\partial h}&=&-\left[\mathcal{R}_h^{(2)}\big(\beta_h(\tau);\tau\big)\right]^{-1}\frac{\partial}{\partial h}\, \mathcal{R}_h^{(1)}\big(\beta_h(\tau);\tau\big)=-\left[\mathcal{R}_h^{(2)}\big(\beta_h(\tau);\tau\big)\right]^{-1}
\frac{1}{h}\,\E\left[X_i\,\tilde{k}\!\left(-\frac{e_i\big(\beta_h(\tau)\big)}{h}\right)\right],
\end{eqnarray*}
with $\tilde{k}(t)=tk(t)$. As in the proof of Lemma 4, $\sup_{(\tau,h)\in[\underline{\tau},\overline{\tau}]\times[\underline{h}_n,\overline{h}_n]} \norm{\widehat{\beta}_h(\tau)-\beta_h(\tau)}=O_p(n^{-1/2})$ implies not only that $\sup_{(\tau,h)\in[\underline{\tau},\overline{\tau}]\times[\underline{h}_n,\overline{h}_n]}\sqrt{\frac{nh}{\ln n}}
\norm{\frac{\partial}{\partial h}\,\widehat{\mathcal{R}}_h^{(1)}\big(\widehat{\beta}_h(\tau);\tau\big)-\frac{\partial}{\partial h}\, \mathcal{R}_h^{(1)}\big(\widehat{\beta}_h(\tau);\tau\big)}=O_p(1)$ but also that $\sup_{(\tau,h)\in[\underline{\tau},\overline{\tau}]\times[\underline{h}_n,\overline{h}_n]}\sqrt{\frac{nh}{\ln n}}
\norm{\frac{\partial\widehat{\beta}_h(\tau)}{\partial h}-\frac{\partial\beta_h(\tau)}{\partial h}}=O_p(1)$. This means that, uniformly in $(h_0,h_1,\tau)$,
\begin{eqnarray*}
\widehat{\beta}_{h_1}(\tau)\!-\!\beta_{h_1}(\tau)-\left[\widehat{\beta}_{h_0}(\tau)\!-\!\beta_{h_0}(\tau)\right]&=&\int_{h_0}^{h_1} \left[\frac{\partial\widehat{\beta}_h(\tau)}{\partial h}-\frac{\partial\beta_h(\tau)}{\partial h}\right]\mathrm{d}h\\
&=&O_p\!\left(\!\sqrt{\frac{\ln n}{n}}\,\right)\int_{h_0}^{h_1}h^{-1/2}\,\mathrm{d}h=O_p\!\left(\!\sqrt{\frac{\ln n}{n}}\,\right)\times \big(\sqrt{h_1}-\sqrt{h_0}\,\big),
\end{eqnarray*}
completing the proof$.\hfill\blacksquare$\vskip 1em

\subsection{The pdf curve and efficient QR estimators}

\textbf{Proof of Proposition 1}~~~~Applying the implicit function theorem to the first-order condition $R^{(1)} (\beta(\tau);\tau)$ yields $\beta^{(1)}(\tau)=D^{-1}(\tau)\E[X]$. It then follows from Lemmata 1 and 4, Assumption~Q2 and (\ref{eq:consistency-uniform}) that
\begin{eqnarray*}
\sup_{\tau,h}\norm{\widehat{R}^{(2)}_h\big(\widehat{\beta}_h(\tau);\tau\big)-R^{(2)}\big(\beta(\tau);\tau\big)}&\le&o(h^s)+O_p\left(\!\sqrt{\frac{\ln n}{n h}} \right)+\sup_{\tau,h}\norm{R^{(2)}\big(\widehat{\beta}_h(\tau);\tau\big)-R^{(2)}\big(\beta(\tau);\tau\big)}\\
&=&o(h^s)+O_p\left(\sqrt{\ln n/(nh)}\right).
\end{eqnarray*}
The result readily follows$.\hfill\blacksquare$\vskip 1em

\noindent\textbf{Proof of Proposition 2}~~~~Let $\mathcal{F}_m$ denote the $\sigma$-field generated by the first $m$ observations and
\begin{displaymath}
\widehat{S}(\tau;n,m,q):=\frac{1}{\sqrt{n-m}}\sum_{i=m+1}^nX_i\,\frac{\mathbb{I}\big[Y_i\le Q(\tau|X_i)\big]-\tau}{q(\tau|X_i)}.
\end{displaymath}
By following the same argument as in \cn{guerre2012uniform}, we find that, conditional on $\mathcal{F}_m$, the Bahadur-Kiefer representations of $\tilde{b}(\tau)$ and $\check{b}(\tau)$ are
\begin{eqnarray*}
\sqrt{n}\Big(\tilde{b}(\tau)-\beta(\tau)\Big)&=&-D_q^{-1}(\tau)\,\widehat{S}(\tau;n,0,q)+O_p\left((\ln n)^{3/4}\,n^{-1/4}\right),\\
\sqrt{n}\Big(\check{b}(\tau)-\beta(\tau)\Big)&=&-D_{\check{q}_h}^{-1}(\tau)\,\widehat{S}(\tau;n,m,\check{q}_h)+O_p\left((\ln n)^{3/4}\,n^{-1/4}\right),
\end{eqnarray*}
using Proposition~3 for $\check{b}(\tau)$ to ensure that the weights $\check{q}_h (\tau|X_i)$ stay, uniformly in $i$, asymptotically bounded away from 0 and infinity. Note that Proposition~3 implies that $\norm{D_{\check{q}_h}^{-1}(\tau)-D_q^{-1}(\tau)}=O_p (\varrho_q)$. As the covariates are bounded by assumption, it follows from Proposition~3 that
\begin{eqnarray*}
\E\left[\left.\left(\widehat{S}(\tau;n,m,\check{q}_h)-\widehat{S}(\tau;n,0,q)\right)^2\right|\mathcal{F}_m\right]&\le&\frac{m}{n}+\E\left[\left. \left(\widehat{S}(\tau;n,m,\check{q}_h)-\sqrt{1-\frac{m}{n}}\,\widehat{S}(\tau;n,m,q)\right)^2\right|\mathcal{F}_m\right]\\
&\le&C\,\frac{m}{n}+2\,\E\left[\left.\left(\widehat{S}(\tau;n,m,\check{q}_h)-\widehat{S}(\tau;n,m,q)\right)^2\right|\mathcal{F}_m\right]\\
&\le&C\left[\frac{m}{n}+\sum_{i=m+1}^n\left(\frac{\check{q}_h(\tau|X_i)-q(\tau|X_i)}{\check{q}_h(\tau|X_i)q(\tau|X_i)}\right)^2\right]\\
&=&O(\varrho_s^2)+O_p(\varrho_q^2).
\end{eqnarray*}
This implies the first part of Proposition~2. The CLT easily follows$.\hfill\blacksquare$\newpage

\begin{table}[th]
\caption{Examples of Gaussian-type kernels}\label{tab:kernel}
\begin{center}\begin{tabular}{lllc}
\multicolumn{4}{p{12cm}}{\small{We provide examples of kernel functions that satisfy Assumption~K1 and their corresponding kernel constants $c_k$. We denote by $\phi$ the standard Gaussian density.}}\\
\hline
Gaussian-type kernel                                                        && kernel constant                    & order ($s+1$)\\
\hline
$k(x)=\phi(x)$                                                              && $c_k=\frac{1}{\sqrt{\pi}}$         & 2\\
$k(x)=\frac{3}{2}\left(1-\frac{x^2}{3}\right)\phi(x)$                       && $c_k=\frac{7}{16\sqrt{\pi}}$       & 4\\
$k(x)=\frac{15}{8}\left(1-\frac{2x^2}{3}+\frac{x^4}{15}\right)\phi(x)$      && $c_k=\frac{321}{1024\sqrt{\pi}}$   & 6\\
$k(x)=\frac{35}{16}\left(1-x^2+\frac{x^4}{5}-\frac{x^6}{105}\right)\phi(x)$ && $c_k=\frac{4175}{16384\sqrt{\pi}}$ & 8\\
\hline
\end{tabular}\end{center}
\end{table}\vfill

\begin{figure}[ht]
\centering
\caption{Examples of quantile paths implied by the standard and smoothed QR estimates\vspace{1em}\newline\small{We consider a sample of 100 observations from $Y=(0.1+0.5X)\epsilon$, where $\epsilon$ is a standard Gaussian error independent of $X\sim U(0,1)$. The left and right panels compare the conditional quantiles of $Y$ respectively given $X=0.1$ and $X=0.9$ (dashed line) with their corresponding standard (grey) and convolution-type kernel (black) QR estimates. We employ the rule-of-thumb bandwidth $h_{ROT}$ for the smoothed QR estimator as in Section 4.}}\label{fig:path}
\vspace{1em}\hspace{-1cm}
\includegraphics{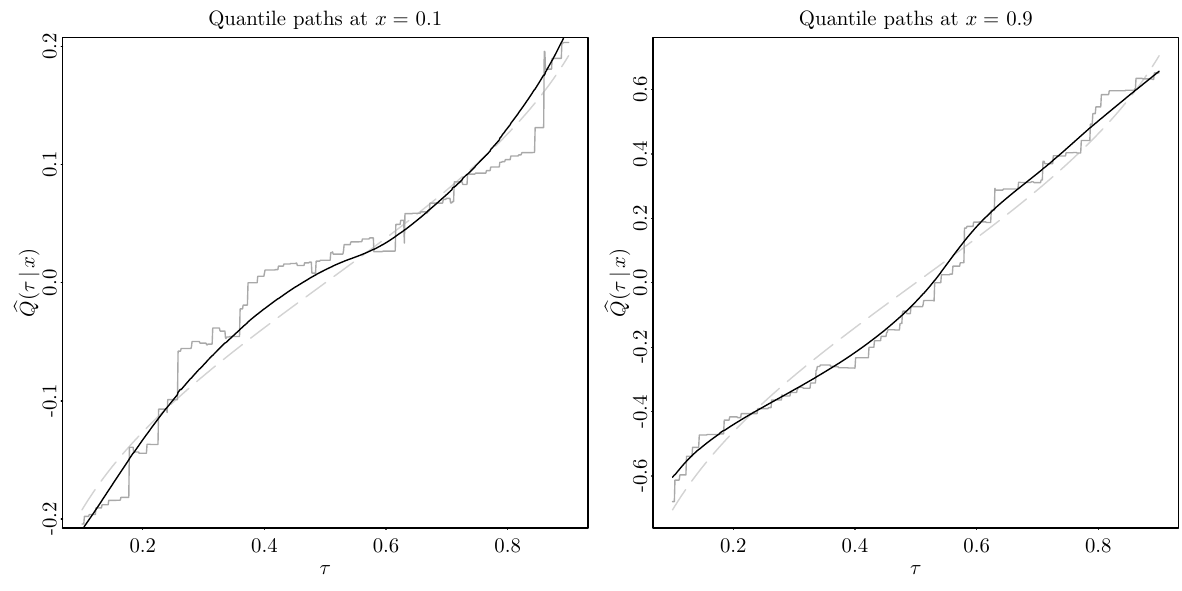}
\end{figure}

\begin{figure}[ht]
\centering
\caption{Conditional quantile function, qdf and pdf estimation examples
\vspace{1em}\newline\small{We sample 200 observations from $Q(\tau|x)=\beta_0(U)+\beta_1(U)X_1+\beta_2(U)X_2+\beta_3(U)X_3$, where $U$, $X_1$, $X_2$ and $X_3$ are independent and uniformly distributed variables on the unit interval $[0,1]$, $\beta_0(\cdot)$ and $\beta_1(\cdot)$ are respectively given by the $\mathsf{Beta}(1,16)$ and $\mathsf{Beta}(32,32)$ quantile functions, $\beta_2(\tau)=1$, and $\beta_3(\tau)=\frac{(2\pi+8)\tau-(\cos(2\pi\tau)-1)}{2\pi+8}$. Panels (a) to (c) are about the conditional quantile function, qdf, and pdf of $Y$ given $x_1=0.5$, $x_2=x_3=0.9$, respectively. True functions are in gray dashed lines, whereas black solid lines correspond to estimates using $h_{ROT}$ as bandwidth.}}\label{fig:pdfonedraw}
\includegraphics[width=\linewidth]{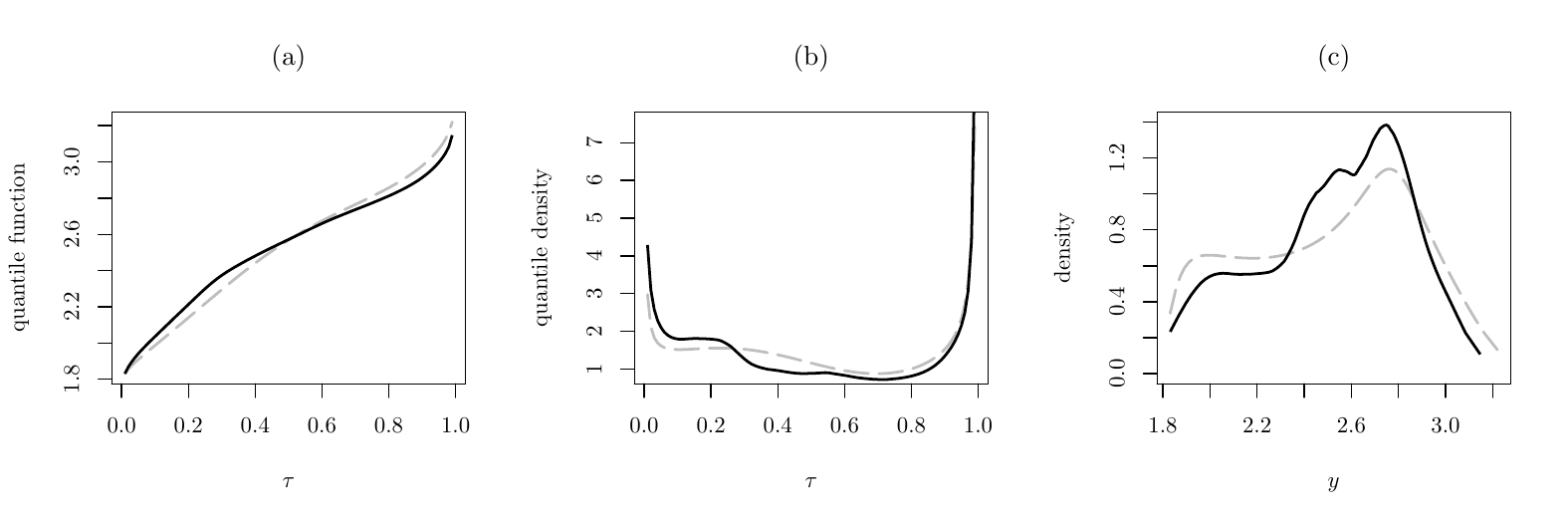}
\end{figure}

\begin{figure}[ht]
\centering
\caption{Conditional quantile function, qdf and pdf estimates\vspace{1em}\newline\small{We consider 1,000 samples of 1,000 observations from $Y=\beta_0(U)+\beta_1(U)X_1+\beta_2(U)X_2+\beta_3(U)X_3$, where $U$, $X_1$, $X_2$ and $X_3$ are independent and uniformly distributed variables on the unit interval $[0,1]$, $\beta_0(\cdot)$ and $\beta_1(\cdot)$ are respectively given by the $\mathsf{Beta}(1,16)$ and $\mathsf{Beta}(32,32)$ quantile functions, $\beta_2(\tau)=1$, and $\beta_3(\tau)=\frac{(2\pi+8)\tau-(\cos(2\pi\tau)-1)}{2\pi+8}$. Panels (a) to (c) are about the conditional quantile function, qdf, and pdf of $Y$ given $(x_1,x_2,x_3)=(0.9,0.5,0.9)$, respectively. Gray shades refer to the [0.05,0.95]-interquantile bands of the corresponding estimators using $h_{ROT}$ as bandwidth, whereas solid black lines depict the true functions.}}\label{fig:pdf-estimator-simu}
\includegraphics{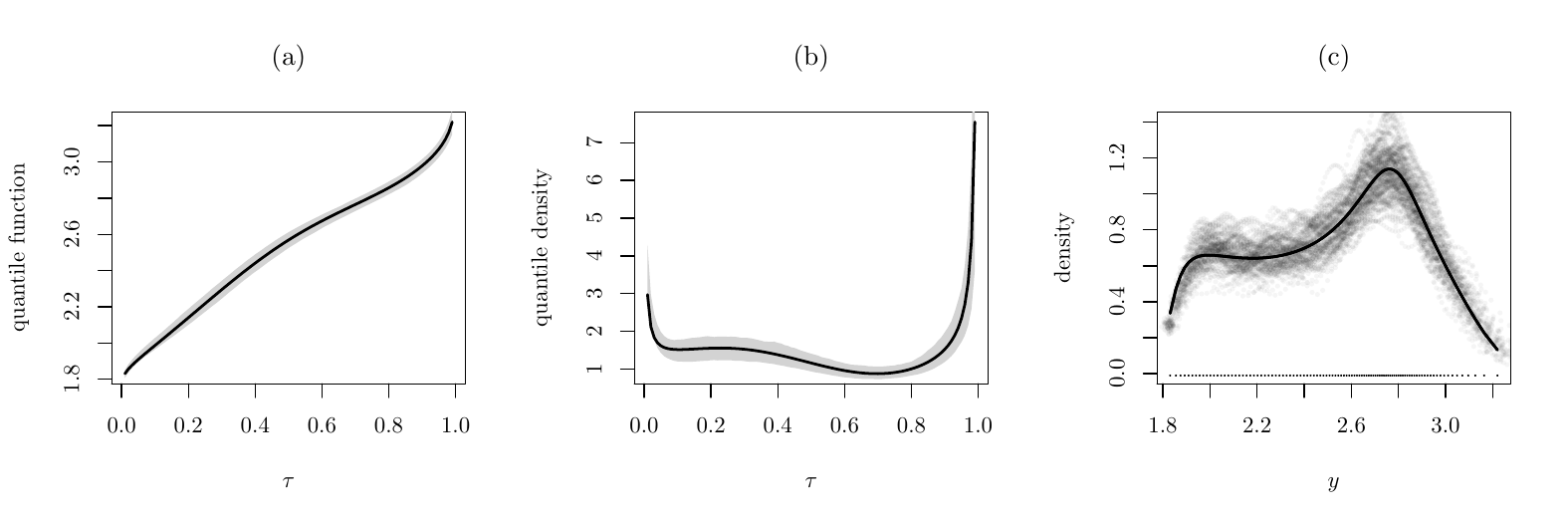}
\end{figure}

\begin{sidewaysfigure}
\centering
\caption{Performance of the MR estimators with exponential errors\vspace{1em}\newline\small{The first row of plots displays the mean squared error relative to the standard MR estimator, whereas the second row depicts the standard error of the slope estimator, within a one-standard-deviation band. The third and fourth rows focus respectively on the empirical coverage of the asymptotic confidence intervals at the 95\% and 99\% levels, as a function of the bandwidth $h$. We consider sample sizes from 100 to 1,000 observations. MR corresponds to Koenker and Bassett's (1978) standard estimator, SMR to Horowitz's (1998) smoothed estimator, and CKMR to our convolution-type kernel estimator. Finally, $h_{ROT}$ denotes the rule-of-thumb bandwidth, whose results we place arbitrarily at its average value across the 100,000 replications.}}\label{fig:exponential}
\vspace{1em}\includegraphics[width=\linewidth]{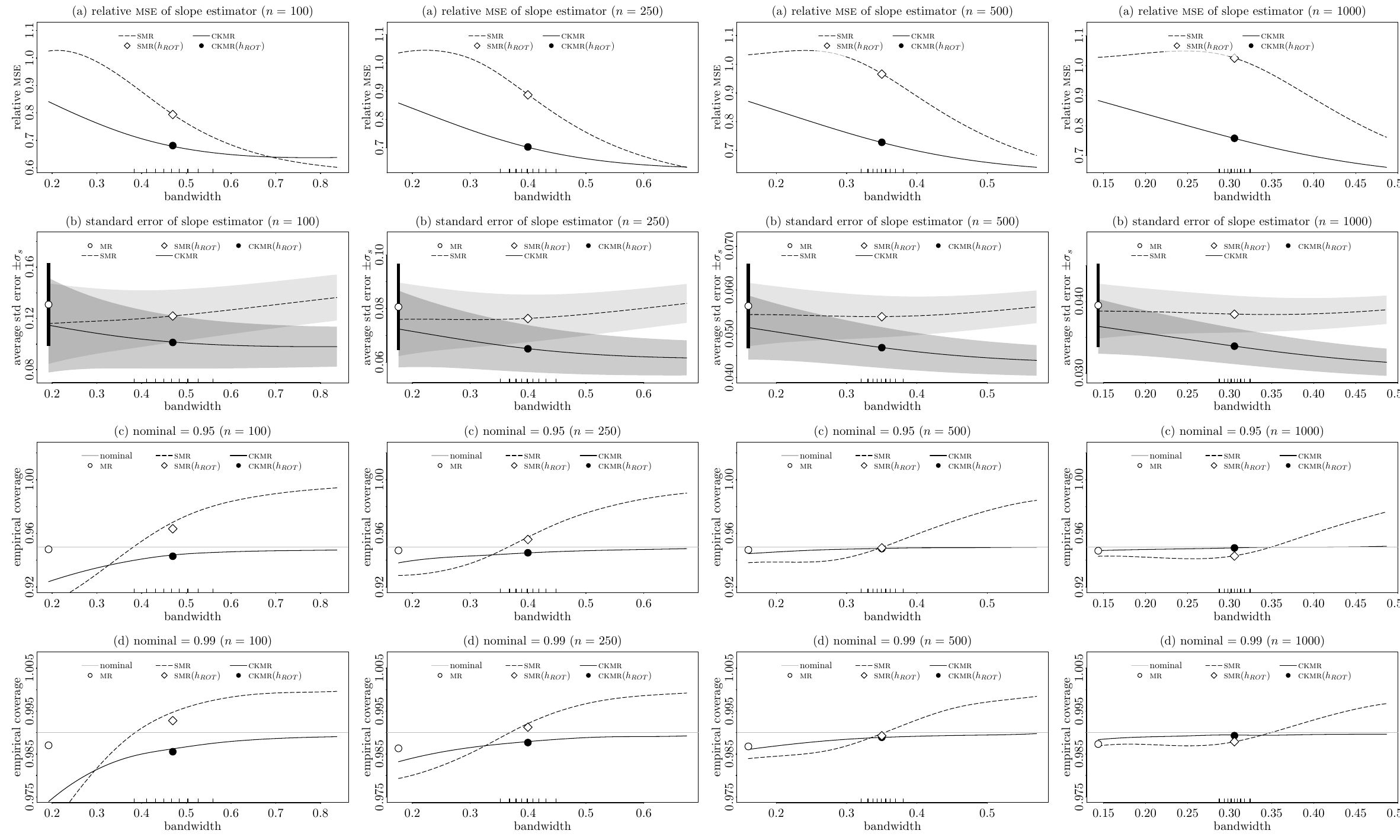}\\
\end{sidewaysfigure}\clearpage

\begin{sidewaysfigure}
\centering
\caption{Performance of the MR estimators with Gumbel errors\vspace{1em}\newline\small{The first row of plots displays the mean squared error relative to the standard MR estimator, whereas the second row depicts the standard error of the slope estimator, within a one-standard-deviation band. The third and fourth rows focus respectively on the empirical coverage of the asymptotic confidence intervals at the 95\% and 99\% levels, as a function of the bandwidth $h$. We consider sample sizes from 100 to 1,000 observations. MR corresponds to Koenker and Bassett's (1978) standard estimator, SMR to Horowitz's (1998) smoothed estimator, and CKMR to our convolution-type kernel estimator. Finally, $h_{ROT}$ denotes the rule-of-thumb bandwidth, whose results we place arbitrarily at its average value across the 100,000 replications.}}\label{fig:gumbel}
	\vspace{1em}\includegraphics[width=\linewidth]{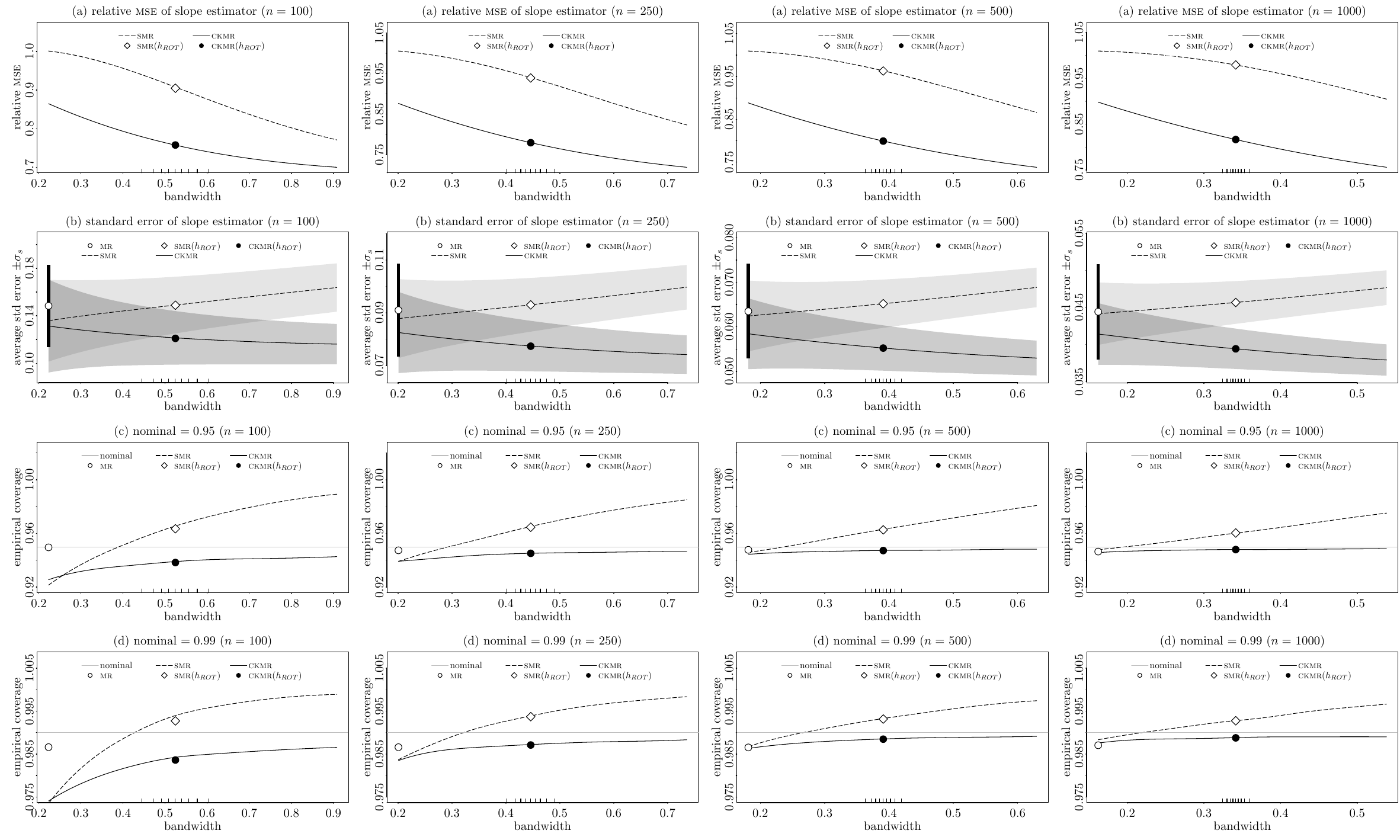}\\
\end{sidewaysfigure}\clearpage

\begin{sidewaysfigure}
\centering
\caption{Performance of the MR estimators with chi-squared errors\vspace{1em}\newline\small{The first row of plots displays the mean squared error relative to the standard MR estimator, whereas the second row depicts the standard error of the slope estimator, within a one-standard-deviation band. The third and fourth rows focus respectively on the empirical coverage of the asymptotic confidence intervals at the 95\% and 99\% levels, as a function of the bandwidth $h$. We consider sample sizes from 100 to 1,000 observations. MR corresponds to Koenker and Bassett's (1978) standard estimator, SMR to Horowitz's (1998) smoothed estimator, and CKMR to our convolution-type kernel estimator. Finally, $h_{ROT}$ denotes the rule-of-thumb bandwidth, whose results we place arbitrarily at its average value across the 100,000 replications.}}\label{fig:chi2}
\vspace{1em}\includegraphics[width=\linewidth]{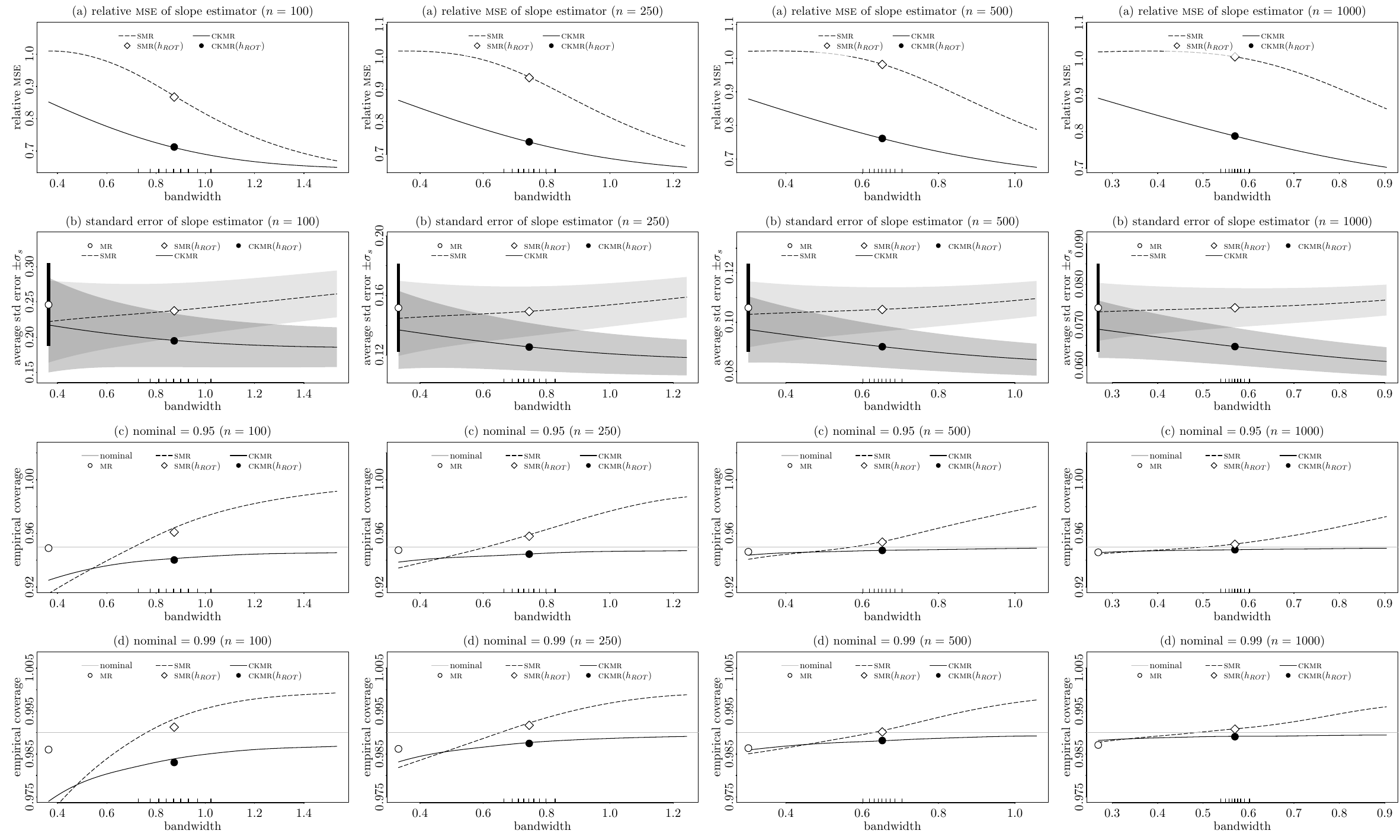}\\
\end{sidewaysfigure}\clearpage

\begin{sidewaysfigure}
\centering
\caption{Performance of the MR estimators with t-student errors\vspace{1em}\newline\small{The first row of plots displays the mean squared error relative to the standard MR estimator, whereas the second row depicts the standard error of the slope estimator, within a one-standard-deviation band. The third and fourth rows focus respectively on the empirical coverage of the asymptotic confidence intervals at the 95\% and 99\% levels, as a function of the bandwidth $h$. We consider sample sizes from 100 to 1,000 observations. MR corresponds to Koenker and Bassett's (1978) standard estimator, SMR to Horowitz's (1998) smoothed estimator, and CKMR to our convolution-type kernel estimator. Finally, $h_{ROT}$ denotes the rule-of-thumb bandwidth, whose results we place arbitrarily at its average value across the 100,000 replications.}}\label{fig:t-student}
\vspace{1em}\includegraphics[width=\linewidth]{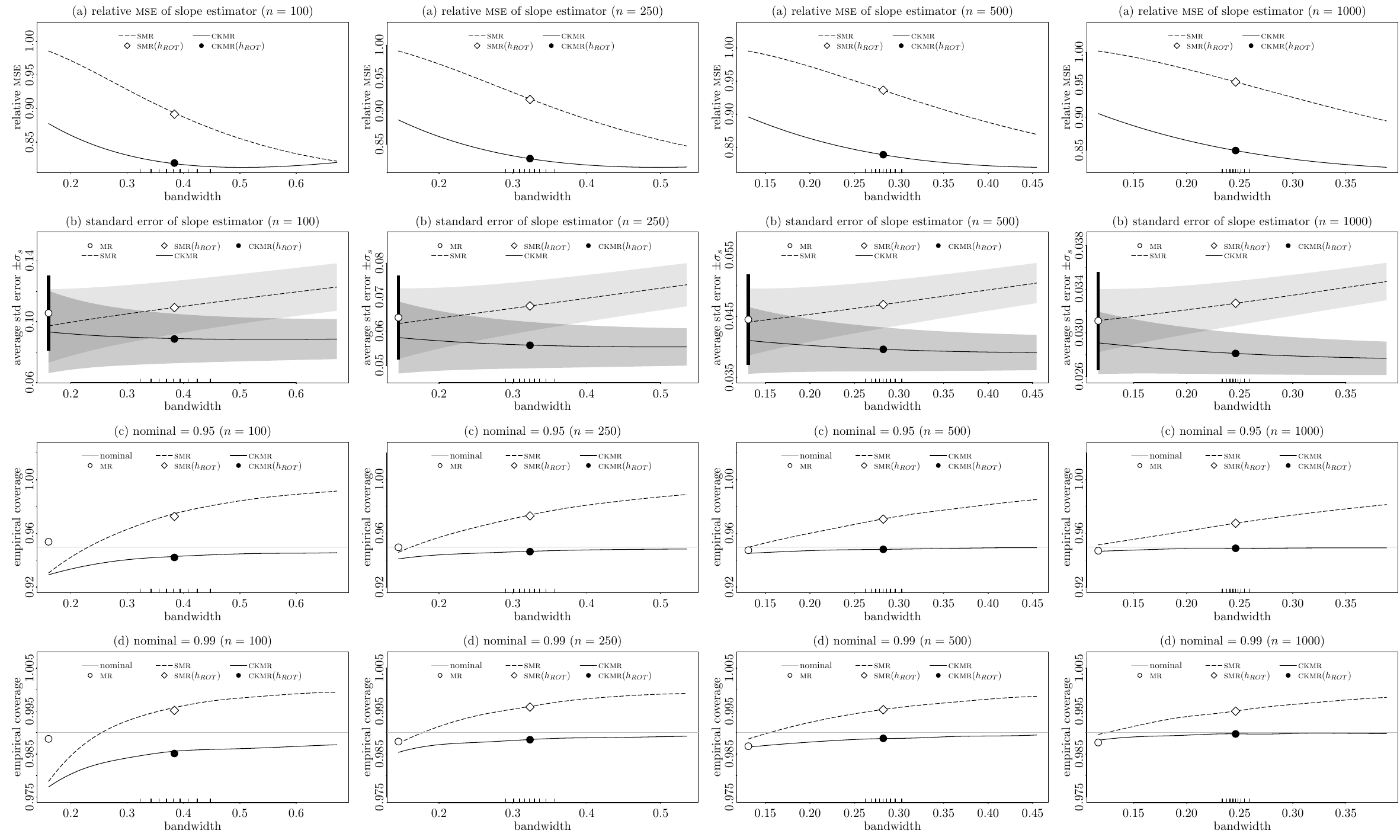}\\
\end{sidewaysfigure}\clearpage

\begin{sidewaysfigure}
\centering
\caption{Performance of the MR estimators with heteroskedastic errors\vspace{1em}\newline\small{The first row of plots displays the mean squared error relative to the standard MR estimator, whereas the second row depicts the standard error of the slope estimator, within a one-standard-deviation band. The third and fourth rows focus respectively on the empirical coverage of the asymptotic confidence intervals at the 95\% and 99\% levels, as a function of the bandwidth $h$. We consider sample sizes from 100 to 1,000 observations. MR corresponds to Koenker and Bassett's (1978) standard estimator, SMR to Horowitz's (1998) smoothed estimator, and CKMR to our convolution-type kernel estimator. Finally, $h_{ROT}$ denotes the rule-of-thumb bandwidth, whose results we place arbitrarily at its average value across the 100,000 replications.}}\label{fig:heteroskedastic}
\vspace{1em}\includegraphics[width=\linewidth]{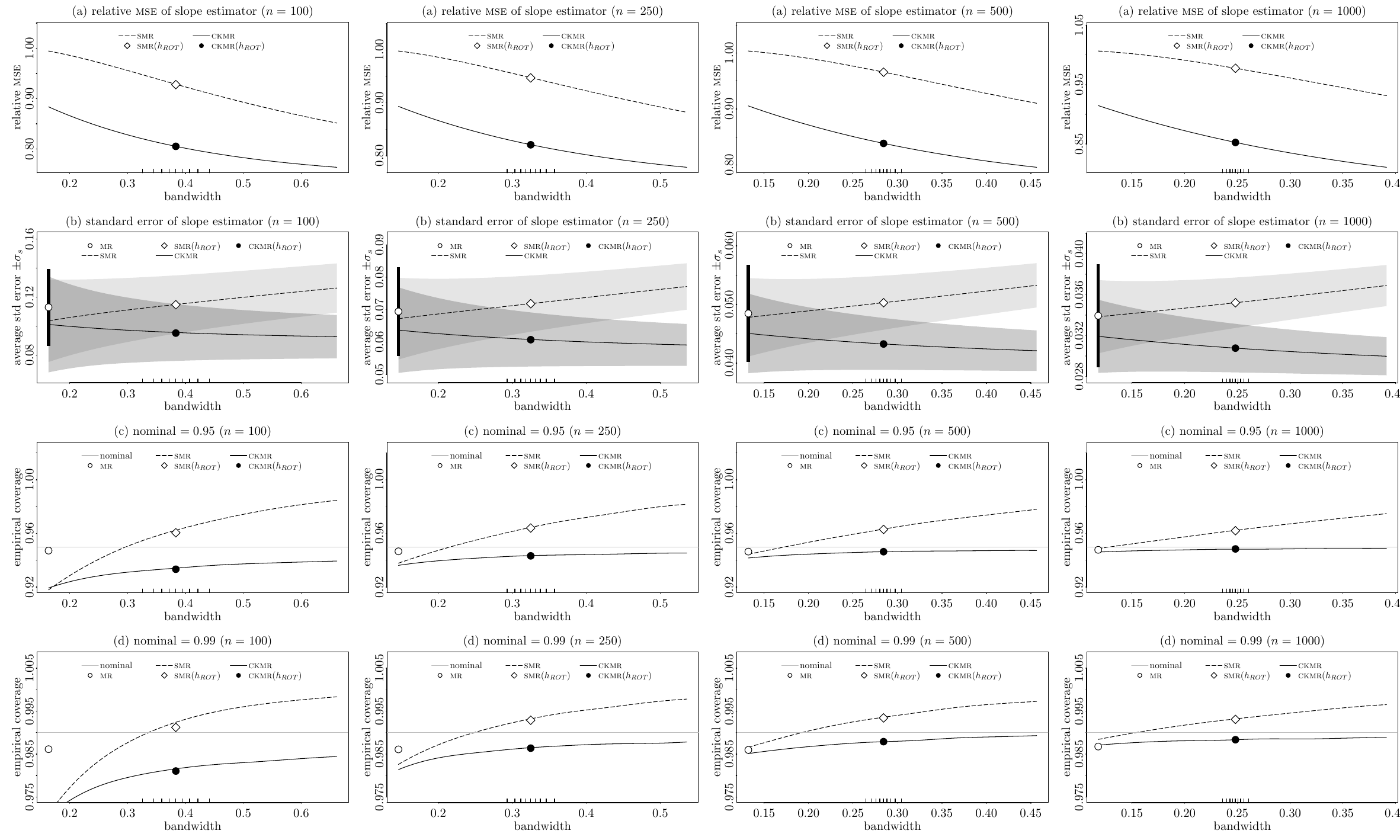}\\
\end{sidewaysfigure}\clearpage

\newpage\lineskip=.35ex \baselineskip 3.5ex
\setcounter{page}{1}\appendix
\begin{center}
{\Large\textsc{Online Appendix}}
\end{center}\vskip 2em

\section{Comparison with Horowitz (1998)}

It is possible to study Horowitz's (1998) smoothed QR estimator using the same tools we employ to document the asymptotic behavior of our convolution-type kernel QR estimator. Let $\tau\in(0,1)$ and Assumptions~X, Q and K hold. Let now $\mathfrak{R}^{(j)}_h(b;\tau):=\E[\widehat{\mathfrak{R}}^{(j)}_h(b;\tau)]$ for $j=0,1,2$ and $\mathfrak{b}_h(\tau):=\arg\min_b\mathfrak{R}_h(b;\tau)$. The latter corresponds to the unique solution of the first-order condition $\mathfrak{R}^{(1)}_h(\mathfrak{b}_h(\tau);\tau)=0$ for $h$ small enough. It turns out that $\widehat{\mathfrak{R}}^{(2)}_h(b;\tau)=\frac{1}{n}\sum_{i=1}^nX_iX_i'\,\kappa\big(e_i(b)/h\big)$, where $\kappa(t):=2k(t)+tk^{(1)}(t)$. Integrating by parts shows that $\int t^jtk^{(1)}(t)\,\mathrm{d}t=-(j+1)\int t^jk(t)\,\mathrm{d}t$, so that $\kappa(\cdot)$ is a kernel function with the same order than $k(\cdot)$. Accordingly, Horowitz's (1998) smoothed objective function also satisfies Lemma 1.

Along the same lines as in the proof of Theorem 1,
\begin{eqnarray*}
\mathfrak{b}_h(\tau)-\beta_h(\tau)&=&-\left[\mathfrak{R}^{(2)}(\beta(\tau);\tau)+o(1)\right]^{-1}\mathfrak{R}^{(1)}_h\big(\beta_h(\tau);\tau\big)\\
                                  &=&\left[D^{-1}(\tau)+o(1)\right]\E\left[X\,\frac{e\big(\beta_h(\tau)\big)}{h}\,k\!\left(-\frac{e\big(\beta_h(\tau)\big)}{h}\right)\right].
\end{eqnarray*}
Because $k(\cdot)$ is symmetric and of order $s+1$, Theorem 1 implies that
\begin{eqnarray*}
\E\left[X\,\frac{e\big(\beta_h(\tau)\big)}{h}\,k\!\left(-\frac{e\big(\beta_h(\tau)\big)}{h}\right)\right]&=&h\,\E\!\left[X\int zk(z)f\big(X'\beta_h(\tau)+hz\given X\big)k(z)\,\mathrm{d}z\right]\\
                                                                                                         &=&h^{s+1}\frac{\int z^{s+1}k(z)\mathrm{d}z}{s!}\,\E\!\left[Xf^{(s)}\big(X'\beta(\tau)\given X\big)\right]+o(h^{s+1})
\end{eqnarray*}
and that $\mathfrak{b}_h(\tau)=\beta_h(\tau)+(s+1)h^{s+1}B(\tau)+o(h^{s+1})=\beta(\tau)+s h^{s+1}B(\tau)+o(h^{s+1})$. This means that $\mathfrak{b}_h(\tau)-\beta(\tau)=-s\big(\beta_h(\tau)-\beta(\tau)\big)+o(h^{s+1})$, so that Horowitz's (1998) smoothing approach amplifies the bias by a factor $-s$ asymptotically.

We next consider the asymptotic covariance matrix of Horowitz's smoothed QR estimator. Consider $b_h(\tau)=\beta(\tau)+O(h^2)$ and let $\Delta_h(\tau):=\widehat{\mathfrak{R}}^{(1)}_h(b_h(\tau);\tau)-\widehat{R}^{(1)}_h(b_h(\tau)b;\tau)$. We first observe that $\mathbb{V}\big[\sqrt{n}\Delta_h(\tau)\big]=O(h)$, whereas using the fact $y=X'b_h(\tau)-hu$ yields under Assumption~Q2 that
\begin{eqnarray*}
n\,\mathrm{Cov}\big(\widehat{R}^{(1)}_h(b_h(\tau);\tau),\Delta_h(\tau)\big)&=&\E\!\left\{XX'\!\!\int\!\left[\tau-K\!\left(\!-\frac{e\big(b_h(\tau)\big)}{h}\!\right)\right]\frac{e\big(b_h(\tau)\big)}{h}\, k\!\left(\!-\frac{e\big(b_h(\tau)\big)}{h}\!\right)f(y|X)\,\mathrm{d}y\right\}\\
&=&h\int[K(u)-\tau]\,uk(u)\,\mathrm{d}u\,\E\left[XX'f\big(X'b_h(\tau)|X\big) \mathrm{d}y\right]+O(h^2)\\
&=&h\int_0^\infty[K(u)-K(-u)]\,uk(u)\,\mathrm{d}u\,\E\left[XX'f\big(X'\beta(\tau)|X\big)\,\mathrm{d}y\right]+O(h^2)
\end{eqnarray*}
for any symmetric kernel $k(\cdot)$. Because $\int_0^\infty[K(u)-K(-u)]\,uk(u)\,\mathrm{d}u>0$ for second-order and \textit{bona fide} higher-order kernels, there exists a symmetric positive $M_\tau$ such that
\begin{displaymath}
\mathbb{V}\left[\sqrt{n}\,\widehat{\mathfrak{R}}^{(1)}_h\big(\beta(\tau);\tau\big)\right]=\mathbb{V}\left[\sqrt{n}\,\widehat{R}^{(1)}_h\big(\beta(\tau);\tau\big)\right]+h\big[M_\tau+o(1)\big].
\end{displaymath}
It then follows from Lemma~1 that $\mathfrak{R}^{(2)}_h\big(\mathfrak{b}_h(\tau);\tau\big)=D(\tau)+o(1)$, and hence
\begin{displaymath}
\mathbb{V}\!\left[\mathfrak{R}^{(2)}_h\big(\mathfrak{b}_h(\tau);\tau\big)^{-1}\,\widehat{\mathfrak{R}}^{(1)}_h\big(\mathfrak{b}_h(\tau);\tau\big)\right]= \mathbb{V}\!\left[R^{(2)}_h\big(\beta_h(\tau);\tau\big)^{-1}\widehat{R}^{(1)}_h\big(\beta_h(\tau);\tau\big)\right]+h\,D^{-1}\!(\tau)M_\tau D^{-1}\!(\tau)+o(h).
\end{displaymath}
Horowitz's estimator has a Bahadur-Kiefer representation as in Theorem 2, ergo the above equality shows that the asymptotic covariance matrix of Horowitz's estimator is larger than ours at the second order.

\section{Technical proofs}

\textbf{Proof of Lemma 1}~~~~Under Assumption~Q2, a Taylor expansion with integral remainder yields
\begin{displaymath}
f(v+hz\given x)=\sum_{\ell=0}^sf^{(\ell)}(v\given x)\,\frac{(hz)^\ell}{\ell!}+\frac{(hz)^{s}}{(s-1)!}\int_0^1(1-w)^{s-1}\left[f^{(s)}(v+whz\given x) -f^{(s)}(v\given x)\right]\mathrm{d}w.
\end{displaymath}

\noindent$\bs{(i)}$~~Assumption~K1 ensures that
\begin{eqnarray}
\E\!\left[k_h(v-Y)\given x\right]-f(v\given x)&=&\int k_h(v-y)\,f(y\given x)\,\mathrm{d}y-f(v\given x)\nonumber\\
&=&\int k(z)\Big[f(v+hz\given x)-f(v\given x)\Big]\,\mathrm{d}z\nonumber\\
&=&\int_0^1(1-w)^{s-1}\!\int\begin{matrix}\frac{(hz)^{s}}{(s-1)!}\end{matrix}\,k(z)\Big[f^{(s)}(v+whz\given x)-f^{(s)}(v\given x)\Big]\,\mathrm{d}z\,\mathrm{d}w\label{eq:conv}
\end{eqnarray}
through a change of variables $y=v+hz$. Now, the check function is such that
\begin{displaymath}
\int\rho_\tau(v)\,\mathrm{d}G(v)=(1-\tau)\int_{-\infty}^0G(v)\,\mathrm{d}v+\tau\int_0^\infty[1-G(v)]\,\mathrm{d}v
\end{displaymath}
for any arbitrary cdf $G$, and hence
\begin{displaymath}
R(b;\tau)=\int\bigg\{(1-\tau)\int_{-\infty}^0\int_{-\infty}^{t+x'b}f(v\given x)\,\mathrm{d}v\,\mathrm{d}t+\tau\int_0^\infty\int_{t+x'b}^\infty f(v\given x)\,\mathrm{d}v\,\mathrm{d}t\bigg\}\,\mathrm{d}F_X(x),
\end{displaymath}
where $F_X(x)$ is the cdf of $X$. Similarly,
\begin{displaymath}
R_h(b;\tau)=\int\bigg\{(1-\tau)\int_{-\infty}^0\int_{-\infty}^{t+x'b}\E\!\left[k_h(v-Y)\given x\right]\mathrm{d}v\,\mathrm{d}t+\tau\int_0^\infty\!\!\int_{t+x'b}^\infty\E\!\left[k_h(v-Y)\given x\right]\mathrm{d}v\,\mathrm{d}t\bigg\}\, \mathrm{d}F_X(x).
\end{displaymath}
It follows from \eqref{eq:conv} that
\begin{eqnarray*}
L_1&:=&\abs{\int_{-\infty}^0\int_{-\infty}^{t+x'b}\bigg\{\E\!\left[k_h(v-Y)\given x\right]-f(v\given x)\bigg\}\mathrm{d}v\,\mathrm{d}t}\\
&=&\abs{\int_0^1(1-w)^{s-1}\int\begin{matrix}\frac{(hz)^{s}}{(s-1)!}\end{matrix}\,k(z)\int_{-\infty}^0\int_{-\infty}^{t+x'b}\left[f^{(s)}(v+whz\given x)-f^{(s)}(v\given x)\right]
	\mathrm{d}v\,\mathrm{d}t\,\mathrm{d}z\,\mathrm{d}w}\\
&=&\abs{\int_0^1(1-w)^{s-1}\int\begin{matrix}\frac{(hz)^{s}}{(s-1)!}\end{matrix}\,k(z)\left[f^{(s-2)}(x'b+whz\given x)-f^{\left(s -2\right)}(x'b\given x)\right]\mathrm{d}z\,\mathrm{d}w}
\end{eqnarray*}
given that $\int\abs{z^{s+1}k(z)}\,\mathrm{d}z<\infty$ by Assumption~K1 and that $f^{(s-2)}(\cdot\given\cdot)$ is Lipschitz. Analogously, $\abs{\int_0^\infty\int_{t+x'b}^\infty\E\!\left[k_h(v-Y)\given x\right]-f(v\given x)\,\mathrm{d}v\,\mathrm{d}t}\le C\,h^{s+1}$, establishing the result.\vskip 1em

\noindent$\bs{(ii)}$~~By the definitions of $R(b;\tau)$ and $R_h(b;\tau)$, it follows from the Lebesgue dominated convergence theorem that
\begin{displaymath}
R^{(1)}(b;\tau)=\E\!\left[X\Big(F(X'b\given X)-\tau\Big)\right]=\int x\bigg[\int_{-\infty}^{x'b}f(y\given x)\,\mathrm{d}y-\tau\bigg]\,\mathrm{d}F_X(x),
\end{displaymath}
and that
\begin{equation}\label{eq:R_h-gradient-representation}
R_h^{(1)}(b;\tau)=\E\!\left\{X\left[K\left(\frac{X'b-Y}{h}\right)-\tau\right]\right\}=\int x\,\bigg\{\int_{-\infty}^{x'b}\E\big[k_h(v-Y)\given x\big]\,\mathrm{d}v-\tau\bigg\}\,\mathrm{d}F_X(x).
\end{equation}
In view that $\int z^sk(z)\,\mathrm{d}z=0$ and $\int\abs{z^{s+1}k(z)}\mathrm{d}z<\infty$, integrating \eqref{eq:conv} yields
\begin{eqnarray}
L_2&:=&\abs{\int_{-\infty}^{x'b}\E\big[k_h(v-Y)\given x\big]-f(v\given x)\,\mathrm{d}v}\nonumber\\
    &=&\abs{\int_0^1(1-w)^{s-1}\!\!\int\begin{matrix}\frac{(hz)^{s}}{(s-1)!}\end{matrix}\,k(z)\int_{-\infty}^{x'b}\left[f^{(s)}(v+whz\given x)-
       f^{(s)}(v\given x)\right]\mathrm{d}v\,\mathrm{d}z\,\mathrm{d}w}\nonumber\\
    &=&\abs{\int_0^1(1-w)^{s-1}\!\!\int\begin{matrix}\frac{(hz)^{s}}{(s-1)!}\end{matrix}\,k(z)\left[f^{(s-1)}(x'b+whz\given x)-
       f^{(s-1)}(x'b\given x)\right]\mathrm{d}z\,\mathrm{d}w}\nonumber\\
    &=&\abs{\int_0^1w(1-w)^{s-1}\!\!\int\begin{matrix}\frac{(hz)^{s+1}}{(s-1)!}\end{matrix}\,k(z)\int_0^1f^{(s)}(x'b+twhz\given x)\mathrm{d}t\,
	   \mathrm{d}z\,\mathrm{d}w}\le C\,h^{s+1}, \label{eq:i1}
\end{eqnarray}
uniformly given that $f^{(s)}$ is bounded. The result then readily follows from Assumption~X.\vskip 1em

\noindent$\bs{(iii)}$~~~~Differentiating $R^{(1)}(b;\tau)$ with respect to $b$ results in
\begin{displaymath}
R^{(2)}(b;\tau)=\E\!\left[XX'f(X'b\given X)\right]=\int xx'f(x'b\given x)\,\mathrm{d}F_X(x)
\end{displaymath}
and, likewise,
\begin{displaymath}
R_h^{(2)}(b;\tau)=\E\!\left[XX'k_h(X'b-Y)\right]=\int xx'\,\E\!\left[k_h(x'b-Y)\given x\right]\mathrm{d}F_X(x).
\end{displaymath}
Setting $v=x'b$ in \eqref{eq:conv} then yields
\begin{eqnarray*}
\norm{R_h^{(2)}(b;\tau)-R^{(2)}(b;\tau)}&\le&C\,\abs{\E\big[k_h(v-Y)\given x\big]-f(v\given x)}\\
                                        &\le&C\,h^s\int\abs{z^s K(z)}\sup_{(x,y)\in\mathbb{R}^{d+1}}\sup_{t:\,\abs{t}\le hz}
                                         \abs{f^{(s)}(y+t\given x)-f^{(s)}(y\given x)}\mathrm{d}z=o(h^s),
\end{eqnarray*}
under Assumptions~X and Q2 by the Lebesgue dominated convergence theorem, as stated.\vskip 1em

\noindent$\bs{(iv)}$~~~~Recall that
\begin{displaymath}
R_h^{(2)}(b;\tau)=\E\!\left[XX'k_h(X'b-Y)\right]=\int k(z)\int xx'f(x'b+hz\given x)\,\mathrm{d}F_X(x)\,\mathrm{d}z.
\end{displaymath}
Under Assumption~Q2, it ensues from $f(\cdot\given\cdot)$ being Lipschitz that
\begin{displaymath}
\norm{R_h^{(2)}(b+\delta;\tau)-R_h^{(2)}(b;\tau)}\le C\int\abs{k(z)}\int\norm{xx'}\abs{x'\delta}\,\mathrm{d}F_X(x)\,\mathrm{d}z\le C\,\norm{\delta},
\end{displaymath}
uniformly in $(b,h,\delta,\tau)$, completing the proof$.\hfill\blacksquare$\vskip 1em

\noindent\textbf{Proof of Lemma 3}~~~~For $\eta>0$,
\begin{eqnarray*}
\left\{\sup_{(\tau,h)}\norm{\widehat{\beta}_h(\tau)-\beta_h(\tau)}\ge 2\eta\right\}&=&\bigcup\limits_{(\tau,h)}\left\{\norm{\widehat{\beta}_h(\tau) -\beta_h(\tau)}\ge 2\eta\right\}\\
&\subset&\bigcup\limits_{(\tau,h)}\left\{\inf_{\left\{b:\norm{b-\beta_h(\tau)}\ge 2\eta\right\}}\widehat{\mathcal{R}}_h(b;\tau)\le \inf_{\left\{b:\norm{b-\beta_h(\tau)}\le 2\eta\right\}} \widehat{\mathcal{R}}_h(b;\tau)\right\}\\
&\subset&\bigcup\limits_{(\tau,h)}\left\{\inf_{\left\{b:\norm{b-\beta_h(\tau)}\ge 2\eta\right\}}\widehat{\mathcal{R}}_h(b;\tau)\le \widehat{\mathcal{R}}_h\big(\beta_h(\tau);\tau\big)\right\}\\
&=&\bigcup\limits_{(\tau,h)}\left\{\inf_{\left\{b:\norm{b-\beta_h(\tau)}\ge 2\eta\right\}}\widehat{\mathcal{R}}_h(b;\tau)\le 0\right\},
\end{eqnarray*}
given that $\widehat{\mathcal{R}}_h\big(\beta_h(\tau);\tau\big)=0$. Theorem~1 ensures that
\begin{eqnarray*}
\Big\{b:\,\norm{b-\beta_h(\tau)}\ge 2\eta\Big\}&\subset&\bigg\{b:\,\norm{b-\beta(\tau)}+\sup_{(\tau,h)}\norm{\beta_h(\tau)-\beta(\tau)}\ge 2\eta\bigg\}\\
&\subset&\Big\{b:\,\norm{b-\beta(\tau)}+O(\bar{h}_n^{s+1})\ge 2\eta\Big\}\\
&\subset&\Big\{b:\,\norm{b-\beta(\tau)}\ge\eta\Big\}
\end{eqnarray*}
for all $(\tau,h)$ provided that $n$ is large enough. This means that
\begin{displaymath}
\left\{\sup_{(\tau,h)}\norm{\widehat{\beta}_h(\tau)-\beta_h(\tau)}\ge 2\eta\right\}\subset\bigcup\limits_{(\tau,h)} \left\{\inf_{\left\{b:\norm{b-\beta(\tau)}\ge\eta\right\}}\widehat{\mathcal{R}}_h(b;\tau)\le 0\right\}.
\end{displaymath}
As $t\mapsto\rho_\tau(t)$ is 1-Lipschitz, it follows from
\begin{displaymath}
\widehat{R}_h(b;\tau)=\frac{1}{nh}\sum_{i=1}^n\int\rho_\tau(t)\,k\!\left(\frac{t-(Y_i-X_i'b)}{h}\right)\mathrm{d}t=\frac{1}{n}\sum_{i=1}^n\int\rho_\tau(Y_i-X_i'b+hz)\,k(z)\,\mathrm{d}z
\end{displaymath}
that
\begin{displaymath}
\abs{\widehat{R}_h(b;\tau)-\widehat{R}(b;\tau)}=\abs{\frac{1}{n}\sum_{i=1}^n\int\Big[\rho_\tau(Y_i-X_i'b+hz)-\rho_\tau(Y_i-X_i'b)\Big]\,k(z)\,\mathrm{d}z}\le h\int\abs{z\,k(z)}\mathrm{d}z<\infty,
\end{displaymath}
for all $b$, $\tau$ and $h$ by Assumption~K1. Theorem~1 and the Lipschitz property of $b\mapsto\widehat{R}(b;\tau)$ then ensures that $\widehat{\mathcal{R}}_h(b;\tau)\ge\widehat{\mathcal{R}}(b;\tau)-C\,h$ uniformly in $b$ and $\tau$, so that
\begin{displaymath}
\left\{\sup_{(\tau,h)}\norm{\widehat{\beta}_h(\tau)-\beta_h(\tau)}\ge 2\eta\right\}\subset\bigcup\limits_{(\tau,h)} \left\{\inf_{\left\{b:\norm{b-\beta(\tau)}\ge\eta\right\}}\widehat{\mathcal{R}}(b;\tau)\le C\,h\right\}.
\end{displaymath}
The next step is a convexity argument. We first perform the change of variables $b=\beta(\tau)+\rho u$ with $\norm{u}=1$ and $\rho\ge\eta$. In view that $b\mapsto\widehat{\mathcal{R}}(b;\tau)$ is convex with $\widehat{\mathcal{R}}\big(\beta(\tau);\tau\big)=0$,
\begin{displaymath}
\frac{\eta}{\rho}\,\widehat{\mathcal{R}}\big(\beta(\tau)+\rho u;\tau\big)=\frac{\eta}{\rho}\,\widehat{\mathcal{R}}\big(\beta(\tau)+\rho u;\tau\big) +\left(1-\frac{\eta}{\rho}\right)\widehat{\mathcal{R}}\big(\beta(\tau);\tau\big)\ge\widehat{\mathcal{R}}\big(\beta(\tau)+\eta u;\tau\big).
\end{displaymath}
It follows from the above inequality that
\begin{displaymath}
\left\{\inf_{\left\{b:\norm{b-\beta(\tau)}\ge\eta\right\}}\widehat{\mathcal{R}}(b;\tau)\le C\,h\right\}\subset\left\{\inf_{\left\{b:\norm{b-\beta(\tau)}=\eta \right\}}\widehat{\mathcal{R}}(b;\tau)\le C\,h\right\},
\end{displaymath}
and hence
\begin{eqnarray*}
\bigcup_{(\tau,h)}\left\{\norm{\widehat{\beta}_h(\tau)-\beta_h(\tau)}\ge 2\eta\right\}&\subset&\bigcup_\tau\left\{\inf_{\left\{b:\norm{b-\beta(\tau)} =\eta\right\}}\widehat{\mathcal{R}}(b;\tau)\le C\,\bar{h}_n\right\}\\
&\subset&\left\{\inf_\tau\inf_{\left\{b:\norm{b-\beta(\tau)}=\eta\right\}}\!\left[\widehat{\mathcal{R}}(b;\tau)\!-\!\mathcal{R}(b;\tau)\right]\le C\,\bar{h}_n\!-\!\inf_\tau\inf_{\left\{b:\norm{b-\beta(\tau)}=\eta\right\}} \!\mathcal{R}(b;\tau)\right\}.
\end{eqnarray*}
We next establish an upper bound for $C\,\bar{h}_n-\inf_{\tau\in[\ubar{\tau},\bar{\tau}]}\inf_{\left\{b:\norm{b-\beta(\tau)}=\eta\right\}}\mathcal{R}(b;\tau)$ using the fact that the eigenvalues of $\mathcal{R}^{(2)}(b;\tau)$ are bounded away from $0$ uniformly in $b$, for $\norm{b-\beta(\tau)}\le 1$ and $\tau\in[\ubar{\tau},\bar{\tau}]$. Given that $R^{(1)}\big(\beta(\tau),\tau\big)=0$, a second-order Taylor expansion of $\mathcal{R}(b;\tau)=R(b;\tau)-R\big(\beta(\tau);\tau\big)$ gives way to
\begin{displaymath}
\mathcal{R}(b;\tau)=0+\big(b-\beta(\tau)\big)'\left[\int_0^1(1-t)\mathcal{R}^{(2)}\big(\beta(\tau)+t\big[b-\beta(\tau)\big];\tau\big)\,\mathrm{d}t\right]
\big(b-\beta(\tau)\big)\ge C\,\eta^2
\end{displaymath}
for all $b$ such that $\norm{b-\beta(\tau)}=\eta$. This means that, for any $\eta_2=\eta-\epsilon_2<\eta$ with conformable $\epsilon_2$ and $\bar{h}_n$ small enough,
\begin{displaymath}
\bigcup_{(\tau,h)}\left\{\norm{\widehat{\beta}_h(\tau)-\beta_h(\tau)}\ge 2\eta\right\}\subset\left\{\sup_{\tau\in[\ubar{\tau},\bar{\tau}]} \sup_{\left\{b:\norm{b-\beta(\tau)}=\eta\right\}}\abs{\widehat{\mathcal{R}}(b;\tau)-\mathcal{R}(b;\tau)}\ge C\,\eta_2^2\right\}.
\end{displaymath}
Now, let $Z_i=(Y_i,X_i')'$, $\theta=(\tau,b')'$ and $g_1(Z_i,\theta)=\rho_\tau(Y_i-X_i'b)-\rho_\tau\big(Y_i-X_i'\beta(\tau)\big)$, so that
\begin{displaymath}
\widehat{\mathcal{R}}(b;\tau)-\mathcal{R}(b;\tau)=\frac{1}{n}\sum_{i=1}^n\left\{g_1(Z_i,\theta)-\E\big[g_1(Z_i,\theta)\big]\right\}.
\end{displaymath}
Under Assumption~X, it follows from $\eta\le 1$ that, for all $b$ such that $\norm{b-\beta(\tau)}=\eta$ and $\tau\in[\ubar{\tau},\bar{\tau}]$,
\begin{displaymath}
\abs{g_1(Z_i,\theta)}\le\norm{X_i}\norm{b-\beta(\tau)}\le C,
\end{displaymath}
implying that $\V\big(g_1(Z_i,\theta)\big)\le\sigma^2\le C$. Observe also that pairing Assumption~X with the Lipschitz conditions on $\tau\mapsto\beta(\tau)$ in Assumption~Q1 and on $\tau\mapsto\rho_\tau(u)$ entails, for all admissible $z$,
\begin{equation}\label{eq:inc}
\abs{g_1(z,\theta_1-g_1(z,\theta_2)}\le C\norm{\theta_1-\theta_2},
\end{equation}
where $\norm{\theta}^2=\norm{b}^2+\abs{\tau}^2$. Next, for $\delta>0$, let $\theta_j$, with $j=1,\dots,J(\delta)\le C\,\delta^{-(d+1)}$, be such that
\begin{displaymath}
\Theta=\Big\{\theta=(b,\tau):\,\tau\in[\ubar{\tau},\bar{\tau}],\norm{b-\beta(\tau)}=\eta_1\Big\}\subset\bigcup_{j=1}^{J(\delta)}\mathcal{B}(\theta_j,\delta),
\end{displaymath}
where $\mathcal{B}(\theta_j,\delta)$ is the $\norm{\cdot}$-ball with center $\theta_j$ and radius $\delta$. Define $\ubar{g}_{1j}(\cdot)$ and $\bar{g}_{1j}(\cdot)$ respectively as $\ubar{g}_{1j}(z):=\inf_{\theta\in\mathcal{B}(\theta_j,\delta)}g_1(z,\theta)$ and $\bar{q}_{1j}(z)=\sup_{\theta\in\mathcal{B}(\theta_j,\delta)}g_1(z,\theta)$, so that $\big\{g_1(\cdot,\theta):\,\theta\in\mathcal{B}(\theta_j,\delta)\big\}\subset[\ubar{g}_{1j},\bar{g}_{1j}]$. Let $\mathcal{G}_{1,\Theta}:=\big\{g_1(\cdot,\theta):\,\theta\in\Theta\big\}\subset\bigcup_{j=1}^{J(\delta)}[\ubar{g}_{1j},\bar{g}_{1j}]$. It follows from \eqref{eq:inc} that $\abs{\bar{g}_{1j}(z)-\ubar{g}_{1j}(z)}\le C\delta\le C$ and $\E\!\left[\abs{\bar{g}_{1j}(Z_i)-\ubar{g}_{1j}(Z_i)}^2\right]\le C\delta^2$. By conditions $(i)$ and $(ii)$ in Lemma~2, it follows from \eqref{eq:maximal-inequality} that setting $H(\delta)=-(d+1)\ln\delta+C$ leads to
\begin{displaymath}
\Prob\left(\sup_{\theta\in\Theta}\abs{\widehat{\mathcal{R}}(b;\tau)-\mathcal{R}(b;\tau)}\ge C\,\frac{1+\sqrt{r}+r/\sqrt{n}}{\sqrt{n}}\right)\le\exp(-r).
\end{displaymath}
This means that, for $n$ large enough with respect to $\eta_2^2$,
\begin{displaymath}
\Prob\left(\sup_\tau\sup_{\big\{b:\,\norm{b-\beta(\tau)}=\eta_1\big\}}\abs{\widehat{\mathcal{R}}(b;\tau)-\mathcal{R}(b;\tau)}\ge C\eta_2^2\right)\le C\exp\left(-n\,C\eta_2^4\right),
\end{displaymath}
and hence
\begin{displaymath}
\Prob\left(\sup_{(\tau,h)}\norm{\widehat{\beta}_h(\tau)-\beta_h(\tau)}\ge 2\eta\right)\le C\exp\left(-n\,C\eta_2^4\right),
\end{displaymath}
completing the proof$.\hfill\blacksquare$\vskip 1em

\noindent\textbf{Proof of Lemma 4}~~~~We start with the first deviation probability. As $R_h^{(1)}\big(\beta_h(\tau),\tau\big)=0$,
\begin{displaymath}
\sup_{(\tau,h)}\norm{\sqrt{n}\,\widehat{R}_h^{(1)}\big[\beta_h(\tau),\tau\big]}\le\sup_{(\tau,h)}\sup_{\big\{b:\,\norm{b-\beta_h(\tau)}\le\eta\big\}} \norm{\sqrt{n}\left(\widehat{R}_h^{(1)}(b,\tau)-R_h^{(1)}(b,\tau)\right)}.
\end{displaymath}
However,
\begin{displaymath}
\widehat{R}_h^{(1)}(b,\tau)=\frac{\partial}{\partial b}\left[\frac{1}{n}\sum_{i=1}^n\int\rho_\tau(Y_i-X_i'b+hz)\,k(z)\,\mathrm{d}z\right]
=\frac{1}{n}\sum_{i=1}^nX_i\left[\int\mathbb{I}(Y_i-X_i'b+hz<0)\,k(z)\,\mathrm{d}z-\tau\right],
\end{displaymath}
implying that $\widehat{R}_h^{(1)}(b,\tau)=\sum_{i=1}^ng_2(Z_i,\theta)/n$, with
\begin{displaymath}
g_2(Z_i,\theta)=X_i\left[\int\mathbb{I}(Y_i-X_i'b+hz<0)\,k(z)\,\mathrm{d}z-\tau\right],
\end{displaymath}
for $Z_i=(Y_i, X_i')'$ and $\theta\in\Theta:=\left\{(b',h,\tau):\,(\tau,h)\in[\ubar{\tau},\bar{\tau}]\times[\ubar{h}_n,\bar{h}_n],\,\norm{b-\beta_h(\tau)} \le\eta\right\}$. We bound each of the entries of $\widehat{R}_h^{(1)}(b,\tau)$, so that there is no loss of generality in assuming that $X_i$ is univariate. Note that $\abs{g_2(Z_i,\theta)}\le C$, $\V\big(g_2(Z_i,\theta)\big)\le\sigma^2\le C$, and $\abs{g_2(Z_i,\theta_2)-g_2(Z_i,\theta_1)}\le C$ for all $\theta_1$ and $\theta_2$. Let $\norm{\theta}^2=\norm{b}^2+\abs{h}^2+\abs{\tau}^2$ and let $\mathcal{B}(\theta,\delta^2)$ denote the $\norm{\cdot}$-ball with center $\theta$ and radius $\delta^2$. Assumption~X ensures that, for any $\theta_1$ and $\theta_2$ in $\mathcal{B}\left(\theta,\delta^2\right)$,
\begin{equation}\label{eq:inc-1}
\abs{g_2(Z_i,\theta_2)-g_2(Z_i,\theta_1)}\le C\left[\int\mathbb{I}\left(Y_i-X_i'b+hz\in[-C\delta^2,C\delta^2]\right)\abs{k(z)}\,\mathrm{d}z+\delta^2\right].
\end{equation}
Consider a covering of $\Theta$ with $J(\delta^2)\le C\,\delta^{-2(d+1)}$ balls $\mathcal{B}(\theta_j,\delta^2)$. Letting $\ubar{g}_{2j}(z):=\inf_{\theta\in\mathcal{B}(\theta_j,\delta)}g_2(z,\theta)$ and $\bar{g}_{2j}(z)=\sup_{\theta\in\mathcal{B}(\theta_j,\delta)}g_2(z,\theta)$ implies not only that $\big\{g_2(\cdot,\theta)\!:\,\theta\in\mathcal{B}(\theta_j,\delta)\big\}\subset[\ubar{g}_{2j},\bar{g}_{2j}]$, but also that $\mathcal{G}_{2,\Theta}:=\big\{g_2(\cdot,\theta)\!:\,\theta\in\Theta\big\}\subset\bigcup_{j=1}^{J(\delta^2)}[\ubar{g}_{2j},\bar{g}_{2j}]$. Equation \eqref{eq:inc-1} ensures that, uniformly in $j$ and $\delta^2\le\sigma^2$,
\begin{displaymath}
\E\!\left[\abs{\bar{g}_{2j}(Z_i)-\ubar{g}_{2j}(Z_i)}^2\right]\le C\,\delta^4+C\,\E\!\left[\int\mathbb{I}\left(Y_i-X_i'b+hz\in[-C\,\delta^2,C\,\delta^2]\right)\abs{k(z)}\mathrm{d}z\right]^2.
\end{displaymath}
Applying the Cauchy-Schwarz inequality under Assumptions~K and Q2 then gives way to
\begin{eqnarray*}
L_4&:=&\E\!\left[\int\mathbb{I}\left(Y_i-X_i'b-hz\in[-C\,\delta^2,C\,\delta^2]\right)k(z)\,\mathrm{d}z\right]^2\\
&\le&\E\!\left[\int\mathbb{I}\left(Y_i-X_i'b-hz\in[-C\,\delta^2,C\,\delta^2]\right)\abs{k(z)}\mathrm{d}z\right]\times\int\abs{k(z)}\mathrm{d}z\\
&\le&\int\E\left\{\Prob\left(Y_i-X_ib-hz\in [-C\,\delta^2,C\,\delta^2]\given X_i\right)\right\}\abs{k(z)}\mathrm{d}z\times\int\abs{k(z)}\mathrm{d}z\\
&\le&C\,\delta^2,
\end{eqnarray*}
implying that $\E\!\left[\abs{\bar{g}_{2j}(Z_i)-\ubar{g}_{2j}(Z_i)}^2\right]\le C(\delta^4+\delta^2)\le C\,\delta^2$, uniformly in $j$ and $\delta^2\le\sigma^2$. As a result, conditions ($i$) and ($ii$) in Lemma~2 hold for $\ln H(\delta)=-2(d+1)\ln\delta+C$, so that \eqref{eq:maximal-inequality} gives
\begin{displaymath}
\Prob\left(\sup_{\theta\in\Theta}\norm{\sqrt{n}\left(\widehat{R}_h^{(1)}(b,\tau)-R_h^{(1)}(b,\tau)\right)}\ge C\,(\sqrt{r}+1+r/\sqrt{n})\right)\le 2\exp(-r).
\end{displaymath}
Accordingly, the first bound holds for $n$ large enough. As for the second bound, there is no loss of generality to assume that $X_i$ is unidimensional. Note that $\sqrt{nh/\ln n}\,\widehat{R}_h^{(2)}(b,\tau) =\sum_{i=1}^ng_3(Z_i,\theta)/\sqrt{n}$, with
\begin{displaymath}
g_3(Z_i,\theta):=\sqrt{\frac{1}{h\ln n}}\,X_i^2\,k\left(\frac{X_i'b-Y_i}{h}\right).
\end{displaymath}
Assumptions~K and X ensure that, uniformly for $\theta\in\Theta$,
\begin{displaymath}
\abs{g_3(Z_i,\theta)}\le C\,\sqrt{\frac{1}{h\ln n}}\le C\,\frac{O(\sqrt{n})}{\ln^2 n}.
\end{displaymath}
It also follows from Assumption~Q2 that, uniformly for $\theta\in\Theta$,
\begin{eqnarray*}
\V\big(g_3(Z_i,\theta)\big)&\le&\frac{C}{h\ln{n}}\int\int k\left(\frac{x'b-y}{h}\right)f(y\given x)\,\mathrm{d}y\,\mathrm{d}F_X(x)\\
&=&\frac{C}{\ln n}\times\int\int k(v)\,f(x'b+hv\given x)\,\mathrm{d}v\,\mathrm{d}F_X(x)\le\frac{C}{\ln n}=\sigma_n^2.
\end{eqnarray*}
Assumption~K posits that, for any $\theta_1$ and $\theta_2$ in $\Theta$, $\abs{g_3(Z_i,\theta_1)-g_3(Z_i,\theta_2)}\le C\,n^C\norm{\theta_1-\theta_2}$. Consider a covering of $\Theta$ with $J(\delta/n^C)\le C\,(\delta/n^C)^{-(d+1)}$ balls $\mathcal{B}(\theta_j,\delta/n^C)$ and let $\ubar{g}_{3j}(z):=\inf_{\theta\in\mathcal{B}(\theta_j,\delta)}g_3(z,\theta)$ and $\bar{g}_{3j}(z):=\sup_{\theta\in\mathcal{B}(\theta_j,\delta)}g_3(z,\theta)$. It then turns out that $\big\{g_3(z,\theta):\theta\in\mathcal{B}(\theta_j,\delta)\big\}\subset[\ubar{g}_{3j},\bar{g}_{3j}]$ and hence $\mathcal{G}_{3,\Theta}=\left\{g_3\left(\cdot,\theta\right):\,\theta\in\Theta\right\}\subset\bigcup_{j=1}^{J(\delta/n^C)}[\ubar{g}_{3j},\bar{g}_{3j}]$, with $\E\!\left[\abs{\bar{g}_3(Z_i)-\ubar{g}_3(Z_i)}^2\right]\le C\,\delta^2$. Conditions ($i$) and ($ii$) in Lemma~2 thus hold for $\ln H(\delta)=-2(d+1)\left(\ln\delta-C\ln n\right)+C$, so that \eqref{eq:maximal-inequality} results for any $u>0$ in
\begin{displaymath}
\Prob\left(\sup_{\theta\in\Theta}\norm{\sqrt{\frac{nh}{\ln n}}\left(\widehat{R}_h^{(2)}(b,\tau)-R_h^{(2)}(b,\tau)\right)}\ge C\left(1+\frac{\sqrt{u}}{\sqrt{\ln n}}+\frac{u}{\ln n}\right)\right)\le 2\exp(-u).
\end{displaymath}
Setting $u=r\ln n$ then yields the exponential inequality.

Suppose now, without loss of generality, that $\mathcal{B}$ is convex. Recall that
\begin{eqnarray*}
\widehat{R}_h^{(2)}(b_1,\tau)-\widehat{R}_h^{(2)}(b_0,\tau)=\frac{1}{n}\sum_{i=1}^nX_iX_i'X_i(b_1-b_0)\int_0^1\frac{1}{h^2}\,k^{(1)}\!\left(\frac{Y_i X_i'\big[b_1+t(b_1-b_0)\big]}{h}\right)\mathrm{d}t
\end{eqnarray*}
and that the variance of $h^{-2}\,k^{(1)}\big((Y_i-X_i'b)/h\big)$ is of order $h^{-3}=o(n/\ln n)$ under Assumption~K. Applying now the same arguments as in the proof of the exponential inequality yields
\begin{eqnarray*}
&&\frac{1}{n}\sum_{i=1}^nX_iX_i'X_i(b_1-b_0)\frac{1}{h^2}\,k^{(1)}\!\left(\frac{Y_i-X_i'\big[b_1+t(b_1-b_0)\big]}{h}\right)\\
&&\quad=\E\left[XX'X(b_1-b_0)\int_{-\infty}^\infty\frac{1}{h}\,k\left(\frac{y-X\big[b_1+t(b_1-b_0)\big]}{h}\right)f^{(1)}(y|X)\mathrm{d}y\right]+O_p\left(\sqrt{\frac{\ln n}{nh^3}}\right),
\end{eqnarray*}
uniformly in $(\tau,h,b_0,b_1)$ for $t\in[0,1]$. The proofs of the remaining results follow similarly$.\hfill\blacksquare$\vskip 1em

\noindent\textbf{Proof of Proposition 3}~~~~Let
\begin{displaymath}
\calEthree:=\left\{\sup_{(\tau,h)}\norm{\widehat{\beta}_h(\tau)-\beta_h(\tau)}\ge\epsilon^{1/4}\right\},
\end{displaymath}
which is such that $\Prob\left(\calEthree\right)\le C\,\exp(-C\,n\,\epsilon)$ by Lemma~3. The bounds for $\Prob\left(\calEone\right)$ and $\Prob\left(\calEtwo\right)$ follow from Lemma~4. In particular, $\lim_{n\rightarrow\infty}\Prob\left(\calEtwo\right)=0$, whereas Lemma~1 ensures under Assumption~X that $b\mapsto\widehat{R}_h(b;\tau)$ is strictly convex for $b$ in a vicinity of $\beta_h(\tau)$, for all $\tau$ in $[\ubar{\tau},\bar{\tau}]$ with probability at least $1-\Prob\left(\calEone\right)-\Prob\left(\calEtwo\right)$. But, by Lemma~3 and Theorem~1, all minimizers of $\widehat{R}_h(b;\tau)$ lie in such a vicinity with a probability tending to $1$. This means that we can make $1-\Prob\left(\calEone\right)-\Prob\left(\calEtwo\right)$ arbitrarily close to $1$ by increasing $r$, and hence $\widehat{\beta}_h(\tau)$ is unique with a probability going to $1$ as $n$ increases. It also follows that, in case $\calEonec$, $\calEtwoc$ and $\calEthreec$ are all true and $n$ is large enough, $\widehat{\beta}_h(\tau)$ satisfies the first-order condition
$\widehat{R}_h^{(1)}\big(\widehat{\beta}_h(\tau);\tau\bigl)=0$. Recall from the proof of Theorem~1 that $\widehat{R}_h^{(2)}(\cdot;\tau)$ has an inverse in the vicinity of $\beta_h(\tau)$ for $n$ large enough on $\mathcal{E}^2(r)$. Applying the implicit function theorem then yields $\widehat{\beta}_h(\tau)$ continuous over the admissible $(\tau,h)$. Accordingly,
\begin{eqnarray*}
-\widehat{R}_h^{(1)}\big(\beta_h(\tau);\tau\big)&=&\widehat{R}_h^{(1)}\big(\widehat{\beta}_h(\tau);\tau\big)-\widehat{R}_h^{(1)}\big(\beta_h(\tau);\tau\big)\\
&=&\big[\widehat{\beta}_h(\tau)-\beta_h(\tau)\big]\int_0^1\widehat{R}_h^{(2)}\left(\beta_h(\tau)+t\big[\widehat{\beta}_h(\tau)-\beta_h(\tau)\big];\tau\right)\mathrm{d}t.
\end{eqnarray*}
Now, if $\epsilon$ in $\calEthree$ is small enough, the eigenvalues of the above matrix are in $\left[1/C,C\right]$ for a large $C$ provided that $n$ is large enough, uniformly in $\tau$ and $h$. This means that
\begin{equation}\label{eq:integral-representation}
\widehat{\beta}_h(\tau)-\beta_h(\tau)=-\left[\int_0^1\widehat{R}_h^{(2)}\left(\beta_h(\tau)+u\big[\widehat{\beta}_h(\tau)-\beta_h(\tau)\big];\tau\right)\,\mathrm{d}u\right]^{-1}\widehat{R}_h^{(1)}\big(\beta_h(\tau);\tau\big).
\end{equation}
Lemma~1($iv$) then implies that, for a generic constant $C$ coming from Bernstein-type inequalities,
\begin{eqnarray*}
P_2&:=&\norm{\sqrt{n}\big(\widehat{\beta}_h(\tau)-\beta_h(\tau)\big)+\big[R_h^{(2)}\!\big(\beta_h(\tau);\tau\big)\big]^{-1}\sqrt{n}\,\widehat{R}_h^{(1)}\!\big(\beta_h(\tau);\tau\big)}\\
&\le&C\norm{\int_0^1\!\left[\widehat{R}_h^{(2)}\!\Big(\beta_h(\tau)+u\big[\widehat{\beta}_h(\tau)\!-\!\beta_h(\tau)\big];\tau\Big)\!-\!R_h^{(2)}\!\Big(\beta_h(\tau)+u\big[\widehat{\beta}_h(\tau)\!-\!\beta_h(\tau)\big];\tau\Big)
	\right]\mathrm{d}u}\norm{\sqrt{n}\,\widehat{R}_h^{(1)}\!\big(\beta_h(\tau);\tau\big)}\\
&&+\;C\norm{\int_0^1\left[R_h^{(2)}\!\left(\beta_h(\tau)+u\big[\widehat{\beta}_h(\tau)-\beta_h(\tau)\big];\tau\right)-R_h^{(2)}\big(\beta_h(\tau);\tau\big)\right]\mathrm{d}u}\norm{\sqrt{n}\,\widehat{R}_h^{(1)}
	\big(\beta_h(\tau);\tau\big)}\\
&\le&C\left\{\sqrt{\frac{\ln n}{nh}}\,r^2+\norm{\widehat{\beta}_h(\tau)-\beta_h(\tau)}\norm{\sqrt{n}\,\widehat{R}_h^{(1)}\big(\beta_h(\tau);\tau\big)}\right\}\\
&\le&C\left\{\sqrt{\frac{\ln n}{nh}}\,r^2+\,n^{-1/2}\norm{\sqrt{n}\,\widehat{R}_h^{(1)}\big(\beta_h(\tau);\tau\big)}^2\right\}\\
&\le&C\left(\sqrt{\frac{\ln n}{nh}}+\frac{1}{\sqrt{n}}\right)r^2
\end{eqnarray*}
on $\calEonec$ and $\calEtwoc$, implying that $\calEc$ holds as long as $C_0$ of the Proposition is large enough$.\hfill\blacksquare$\vskip 1em

\noindent\textbf{Proof of Lemma 5}~~~~Let $h=h_n$ to simplify notation. We first note that $\E\big(\sqrt{n}\,\widehat{S}_h(\tau)\big)=0$. In addition, for any $\alpha,\tau\in [\ubar{\tau},\bar{\tau}]$, it follows that
\begin{displaymath}
\V\big(\sqrt{n}\,\widehat{S}_h(\tau),\sqrt{n}\,\widehat{S}_h(\varsigma)\big)=\E\left\{XX'\left[K\!\left(-\frac{e\big(\beta_h(\tau)\big)}{h}\right)-\tau\right] \left[K\!\left(-\frac{e\big(\beta_h(\varsigma)\big)}{h}\right)-\varsigma\right]\right\}
\end{displaymath}
converges to $\E\!\left\{XX'\big(\mathbb{I}[X'\beta(\tau)\ge Y]-\tau\big)\big(\mathbb{I}[X'\beta(\varsigma)\ge Y]-\varsigma\big)\right\}$ as $n\to\infty$. A simple computation using iterated expectations then yields the limiting covariance structure in \eqref{eq:asymptotic-gaussian-covariance}.

By the Cram\'{e}r-Wold device, in order to obtain weak convergence for the $d$-dimensional process $\big\{\sqrt{n}\,\widehat{S}_h:\,\tau\in[\ubar{\tau},\bar{\tau}]\big\}$, it suffices to consider the convergence in distribution of the linear form $\big\{\sqrt{n}\,\lambda'\widehat{S}_h:\,\tau\in[\ubar{\tau},\bar{\tau}]\big\}$, where $\lambda$ is an arbitrary (fixed) vector in $\mathbb{R}^d$. Assume without loss of generality that $\Vert X\Vert\le 1$ and $\Vert\lambda\Vert\le 1$, and let $Z=(Y,X)\in\mathbb{R}\times\mathbb{R}^d$ and, similarly, $Z_i=(Y_i,X_i)$. Define now $g_{n,\tau}:\mathbb{R}\times\supp X\to\mathbb{R}$ for $z=(y,x)$ as
\begin{equation}\label{eq:Donsker-class}
g_{n,\tau}(z):=x_\lambda\left\{K\left(\frac{x'\beta_h(\tau)-y}{h}\right)-\tau\right\},
\end{equation}
where $x_\lambda=\lambda'x$ and $X_\lambda=\lambda'X$, and consider the class of functions $\mathcal{G}_n=\{g_{n,\tau}:\,\tau\in[\ubar{\tau},\bar{\tau}]\}$. Letting $\mathbb{P}$ and $\mathbb{P}_n$ respectively denote the distribution of $Z$ and the empirical distribution of the sample $(Z_1,\ldots,Z_n)$ yields
\begin{displaymath}
\sqrt{n}\,\lambda'\widehat{S}_h(\tau)=\sqrt{n}\big(\mathbb{P}_n g_{n,\tau}-\mathbb{P}g_{n,\tau}\big).
\end{displaymath}
In other words, the process $\big\{\sqrt{n}\lambda'\widehat{S}_h:\,\tau\in[\ubar{\tau},\bar{\tau}]\big\}$ is an empirical process indexed by a (changing) class of functions $\mathcal{G}_n$. By Theorem~19.28 in \cn{van1998}, it suffices to establish that
\begin{equation}\label{eq:L2-uniform-convergence}
\sup_{\vert\tau-\varsigma\vert<\delta(n)}\E\big\vert g_{n,\tau}(Z)-g_{n,\varsigma}(Z)\big\vert^2\to 0
\end{equation}
and that, for any $\delta(n)\downarrow 0$,
\begin{equation}\label{eq:bracketing-entropy-integral}
\int_0^{\delta(n)}\sqrt{\ln N_{[\,]}(\epsilon,\mathcal{G}_n,L^2(\mathbb{P}))}\,\mathrm{d}\epsilon\to 0
\end{equation}
with $N_{[\,]}(\epsilon,\mathcal{G}_n,L^2(\mathbb{P}))$ denoting the minimum number of $\epsilon$-brackets in $L^2(\mathbb{P})$ required to cover $\mathcal{G}_n$. The remaining requirements of Theorem~19.28 indeed hold trivially in view that the index set $[\ubar{\tau},\bar{\tau}]$ is a compact---and so, totally bounded---metric space, and that the changing classes $\mathcal{G}_n$ admit envelope functions $G_n\equiv 1$ for all $n$ that satisfy the Lindeberg condition $\E_\mathbb{P}\big(G_n^2\,\mathbb{I}[G_n>\sqrt{n}\epsilon]\big)\to 0$.

Let $\partial_\tau:=\frac{\partial}{\partial\tau}$. By Lemma~1 and Theorem~1, applying twice the implicit function theorem yields
\begin{displaymath}
\partial_\tau\beta_h(\tau)=-D_h(\tau)^{-1}\,\partial_\tau R_h^{(1)}\big(\beta_h(\tau);\tau\big)=D_h(\tau)^{-1}\E(X)=\left[D(\tau)+o(1)\right]^{-1}\E(X)=\partial_\tau\beta(\tau)+o(1)
\end{displaymath}
uniformly for $(\tau,h)\in[\ubar{\tau},\bar{\tau}]\times[\ubar{h},\bar{h}]$. This implies, by Assumption~Q1, that $\sup\norm{\partial_\tau\beta_h(\tau)}\le C$ for $n$ large enough, with supremum taken over $(\tau,h)\in[\underline{\tau},\overline{\tau}]\times[\underline{h}_n,\overline{h}_n]$, and so $\Vert\beta_h(\tau)-\beta_h(\varsigma)\Vert\le C\vert\tau-\varsigma\vert$. It also follows from the inverse function theorem and Assumption~Q1 that $\tau\mapsto x'\beta_h(\tau)$ is strictly increasing in $\tau$, for any $x\in\supp X$ and $n$ large enough. In what follows, we assume that $n$ is large enough, so that the above holds.

Now, let $\ubar{\tau}\le\tau_\mathrm{L}\le\tau_\mathrm{U}\le\bar{\tau}$ and consider two random elements (possibly degenerate) $\widehat{\tau}$ and $\widehat{\varsigma}$ in $[\tau_\mathrm{L},\tau_\mathrm{U}]$. The mean value theorem and Assumption~Q2 then ensure that
\begin{equation}\label{eq:crucial-bound}
\Pr\Big(x'\beta_h(\widehat{\tau}\wedge\widehat{\varsigma})-hu\le Y\le x'\beta_h(\widehat{\tau}\vee\widehat{\varsigma})-hu\given X=x\Big)\le C\abs{\tau_\mathrm{U}-\tau_\mathrm{L}},
\end{equation}
uniformly for $u\in\mathbb{R}$ and $x\in\supp X$, given that $\big[x'\beta_h(\widehat{\tau}\wedge\widehat{\varsigma}),\,x'\beta_h(\widehat{\tau} \vee\widehat{\varsigma})\big]\subset\big[x'\beta_h(\tau_\mathrm{L}),\,x'\beta_h(\tau_\mathrm{U})\big]$ and $\abs{x'\beta_h(\tau_\mathrm{U})-x'\beta_h(\tau_\mathrm{L})}\le C\abs{\tau_\mathrm{U}-\tau_\mathrm{L}}$. Define $\Upsilon_u=\{X'\beta_h(\widehat{\tau}\wedge\widehat{\varsigma})-Y\le hu\le X'\beta_h(\widehat{\tau}\vee\widehat{\varsigma})-Y\}$. It follows from $\big\vert g_{n,\widehat{\tau}}(Z)-g_{n,\widehat{\varsigma}}(Z)\big\vert\le\int\mathbb{I}(\Upsilon_u)\abs{k(u)}\mathrm{d}u+\abs{\widehat{\tau}-\widehat{\varsigma}}$ that
\begin{align}
\E\big\vert g_{n,\widehat{\tau}}(Z)-g_{n,\widehat{\varsigma}}(Z)\big\vert^2&\le\E\vert\widehat{\tau}-\widehat{\varsigma}\vert^2+2\E\Big[\abs{\widehat{\tau}-\widehat{\varsigma}} \int\mathbb{I}(\Upsilon_u)\abs{k(u)}\mathrm{d}u\Big]+\E\Big[\int\mathbb{I}(\Upsilon_u)\abs{k(u)}\mathrm{d}u\Big]^2\nonumber\\
&\le\abs{\tau_\mathrm{U}-\tau_\mathrm{L}}^2+2\abs{\tau_\mathrm{U}-\tau_\mathrm{L}}\int\Pr(\Upsilon_u)\abs{k(u)}\mathrm{d}u+C\int\Pr(\Upsilon_u)\abs{k(u)}\mathrm{d}u\label{eq:L2-norm-upper-bound}\\
&\le C\abs{\tau_\mathrm{U}-\tau_\mathrm{L}},\label{eq:L2-bound}
\end{align}
given that the Cauchy-Schwarz inequality implies that $\int\mathbb{I}(\Upsilon_u)\abs{k(u)}\mathrm{d}u\int\abs{k(u)}\mathrm{d}u$ is an upper bound for $\left(\int\mathbb{I}(\Upsilon_u)\abs{k(u)}^{1/2}\abs{k(u)}^{1/2}\,\mathrm{d}u\right)^2$, $\int\abs{k(u)}\mathrm{d}u<\infty$ by Assumption~K1, and $\Pr(\Upsilon_u)=\E\big[\Pr(\Upsilon_u\given X)\big]\le C\abs{\tau_\mathrm{U}-\tau_\mathrm{L}}$ by iterated expectations and \eqref{eq:crucial-bound}. Taking $\widehat{\tau}$ and $\widehat{\varsigma}$ to be deterministic shows that \eqref{eq:L2-uniform-convergence} holds, for all $\delta(n)\downarrow 0$.

We now obtain a set of brackets whose bracketing number is of order $1/\epsilon$. For $\epsilon>0$ small enough, we cover the interval $[\ubar{\tau},\bar{\tau}]$ with $J(\epsilon)\le\lceil(\bar{\tau}-\ubar{\tau})/\epsilon+1\rceil\le 2/\epsilon$ open intervals $B_i=(\tau_i-\epsilon,\tau_i+\epsilon)$, and let $\bar{g}_n^i(z)=\sup_{\tau\in B_i}g_{n,\tau}(z)$ and $\ubar{g}_n^i(z)=\inf_{\tau\in B_i}g_{n,\tau}(z)$. It is straightforward to appreciate that the collection formed by the brackets $[\ubar{g}_n^i,\,\bar{g}_n^i]$, with $i=1,\dots,J(\epsilon)$, covers $\mathcal{G}_n$ and that these suprema and infima are attained in the closure of $B_i$.\footnote{~~For $i=1$ and $i=J(\epsilon)$, the intervals are actually $[\ubar{\tau},\tau_1+\epsilon]$ and $[\tau_{J(\epsilon)}-\epsilon,\bar{\tau}]$, respectively. For simplicity of exposition, we keep the notation as above.} In particular, $\bar{g}_n^i(Z)=g_{n,\widehat{\tau}_i}(Z)$ and $\ubar{g}_n^i(Z)=g_{n,\widehat{\varsigma}_i}(Z)$, where $\widehat{\tau}_i$ and $\widehat{\varsigma}_i$ are random elements in $[\tau_i-\epsilon,\tau_i+\epsilon]$. Resorting to \eqref{eq:L2-bound} once more then gives
\begin{displaymath}
\E\vert\bar{g}_n^i(Z)-\ubar{g}_n^i(Z)\vert^2\le C\epsilon,
\end{displaymath}
and, as a result, $N_{[\,]}(\epsilon,\mathcal{G}_n,L^2(P))\le C/\epsilon$. This ensures that \eqref{eq:bracketing-entropy-integral} holds for all $\delta(n)\downarrow 0.\hfill\blacksquare$

\end{document}